\documentclass[prd,showpacs,floatfix,reprint,preprintnumbers,amsmath,amssymb,superscriptaddress,nofootinbib]{revtex4}

\newcommand{\ba}{\nopagebreak[3]\begin{eqnarray}}
\newcommand{\ea}{\end{eqnarray}}
\usepackage{graphicx}
\usepackage{dcolumn}
\usepackage{bm}
\usepackage{ulem}
\usepackage{xcolor}
\newcommand{\be}{\nopagebreak[3]\begin{equation}}
\newcommand{\ee}{\end{equation}}

\begin{document}


\title{Cosmological constraints on  unimodular gravity models with diffusion }


\author{Susana J. Landau}%
 \email{slandau@df.uba.ar}
\affiliation{Instituto de  F\'{\i}sica de Buenos Aires - CONICET - Universidad de Buenos Aires, Argentina}
\affiliation{CONICET, Godoy Cruz 2290, 1425 Ciudad Aut\'{o}noma de Buenos Aires, Argentina} 

\author{Micol Benetti}
 \affiliation{Scuola Superiore Meridionale, Universit\`{a} di Napoli ``Federico II'',  Largo San Marcellino 10, 80138 Napoli, Italy}
  \affiliation{Dipartimento di Fisica ‘E. Pancini’, Universit\'a di Napoli ‘Federico II’, Compl.
Univ. di Monte S. Angelo, Edificio G, Via Cinthia, I-80126, Napoli, Italy }
 \affiliation{Instituto Nazionale di Fisica Nucleare (INFN), Sez. di Napoli, Complesso Univ. Monte S. Angelo, Via Cinthia 9, 80126, Napoli, Italy}

\author{Alejandro Perez}
\affiliation{Aix Marseille Univ, Université de Toulon, CNRS, CPT, France}%

\author{Daniel Sudarsky}
\affiliation{Instituto de Ciencias Nucleares - Universidad Nacional Aut\'{o}noma de M\'{e}xico, Mexico}%


\date{\today}

\begin{abstract}

A discrete space-time  structure  lying at  about the Planck scale   may become manifest
in the form of  very small violations of the conservation of the matter energy-momentum tensor.  In order to include such kind of violations,  forbidden 
  within the  General Relativity  framework, the theory of   unimodular gravity   seems  as the simplest option to describe the gravitational interaction.
   In the cosmological context, 
  a  direct consequence of  such  violation of  energy conservation  might be  heuristically   viewed  a  ``diffusion process of matter 
  (both dark and ordinary)” into  an effective  dark  energy term in Einstein's equations, which  leads  under   natural    assumptions to  an  adequate estimate for the value of the cosmological constant.  
  Previous works have  also indicated  that these kind of models  might offer a natural scenario to  alleviate the Hubble tension. 
  In this work, we consider a simple  model for the
cosmological history including  a late time   occurrence  of such  energy violation and study   the  modifications of  the predictions for the anisotropy and polarization of the 
Cosmic Microwave Background (CMB).    We  compare the model's predictions with recent data from the CMB, 
Supernovae Type Ia, cosmic chronometers and Baryon Acoustic Oscillations. The results show the potential of this type of  model to alleviate the Hubble tension.
\end{abstract}

\maketitle


\section{Introduction}

 Modern cosmology  relies on  Einstein's equation, and  thus,  on the  strict  conservation of energy momentum  tensor. This is a feature that  seems  essentially   untouchable.
 However a deeper  look  at the issue  reveals that  conservation of energy momentum    is  unlikely to  be  an exact feature of nature \cite{Maudlin:2019bje}.
When taking into account  the quantum  nature  of matter,  it  becomes  rather apparent  that  our gravitational   theory must   also undergo  some  modifications
which are likely to include at least  small departures from  exact conservation laws. However, these changes  might be larger   in early cosmological times and could have cumulative  effects  with  empirically relevant consequences.

Recently,  a group involving some of  us  \cite{Josset:2016vrq},  used of the fact 
 that in  Unimodular Gravity  (UG)  the  strict conservation of  the  energy momentum  tensor is  not required for the   consistency of the theory, in contrast   with what  occurs in General Relativity (GR).
 The theory might be said to reduce the  general invariance   under  diffeomorphisms, which is characteristic of GR to  a more restricted invariance under four-volume  preserving  diffeomorphisims\footnote{Strictly speaking   the invariance under diffeomorphisms  is tautological  in any theory that relies on   manifold  and tensor fields.  The  point, as discussed  in   \cite{Wald:1984}
     is   whether or not  all geometrical  entities entering the theory  are  related (or  derived  from)   to  the fundamental   degrees  of freedom,   which in  theories of  gravitation correspond to the   space-time metric. When  some   additional  non-dynamical   structure is attached to   space-time and left untouched by the diffeomorphic transformation is  that  one finds what  are called  violations  of full diffeomorphism   invariance.  In the case under consideration such structures could be  either the violations  intrinsically   associated  with resolving the    conceptual problems of quentum theory as  discusssed in 
     \cite{Maudlin:2019bje},  or the structure characterizing the  discreteness of spacetime.  }.
Concretely  \cite{Josset:2016vrq}  studied  within that scheme---i.e. violation of energy momentum conservation  together with unimodular gravity---the idea that a  violation of the  conservation of the energy momentum tensor might  have  important and  observable consequences on the    value of the effective  cosmological constant.  Of course, the idea under consideration  contemplates a   violation that is too small to be directly detectable, as no  direct observation of  any departure from  strict  energy  conservation  has  been  observed  at this stage.  Further studies \cite{Perez:2018wlo, SP2019},  considered the idea  that  such violations 
 might   also result from defects in the fabric  of  space-time, and its 
 ultimate origin might lie at the level of the  quantum gravitational description thereof. 
 That  analysis resulted  on  an  attractive account for the   nature  and   magnitude of the   cosmological constant \cite{Perez:2018wlo, SP2019}, which   without the use of   any kind  of fine tuning, leads  to  an estimate for the amount of the  dark energy  which  agrees with those  emerging  from astronomical  and  cosmological observations. In that  proposal an effective cosmological constant represents the  accumulated effect of    extremely   small  departures  of  such  conservation  with the dominant part of the effect arising  around the electro-weak  scale (yielding a  value  $ \Lambda \sim m_p^2 ( E_{ew}/m_p)^7 $ which  can easily  be seen  to be  of correct order of magnitude).  The perspective provided by this work motivated the investigation of the possible role of the fundamental granularity of Planckian physics as the primordial cause of inhomogeneities ultimately observed at the CMB. This led to an alternative paradigm to the standard inflationary one where inhomogeneities at the CMB are directly linked with Planckian discreteness during a primordial inflationary era \cite{Amadei:2021aqd}.

In some recent  work
  \cite{Perez:2019gyd,Perez:2020cwa} the issue of the so-called $H_0$ tension---i.e. the incompatibility at more than $4 \sigma$ between the  value of $H_0$ obtained in  the   context of   the standard cosmological model  by the use of   latest CMB data \cite{planckcosmoparams18} and what is obtained  in  studies  based on   type Ia supernovae (SnIa) explosions together with local distance calibrations (CC)  \cite{2019ApJ...876...85R,2021arXiv211204510R}---has been considered from the present perspective. 
In the proposal for the  generation of  a cosmological constant in the context of UG gravity, the conditions that resulted in  the relatively large violations of the energy-momentum tensor conservation  are tied to regimes  in which  large space-time curvatures are involved. The proposal  to deal  with the $H_0$ tension is  meant to be  an extension  of  such account, which  thus must involve situations  where    similar high curvature conditions prevail in the  post recombination   epoch (where the  effective  cosmological constant  would  have to change by  a nontrivial  amount). At late   cosmological times  the  only situations where such strong gravitational fields can  be reasonably    assumed to arise  correspond to the  late time  dynamics  of  black holes,  particularly  highly rotating ones,  which  might lose up to  $1/3$  of their  mass if slowed down via a diffusion mechanism without  violating the  second  law of thermodynamics\footnote{Which  we   will take  to  continue to hold even in those  extreme   circumstances.}.

Such preliminary study, which assumed a simple expression for the violation of the energy-momentum tensor, concluded that the overall plausibility of the model appears to be supported by observational data \cite{Perez:2020cwa}. 
That analysis focused on  the effects on  the value of the acoustic angular scale of the CMB and   the comparison  with the corresponding  recent observational data, and therefore was able to consider  just  the changes in the  global  geometry of the universe introduced by the UG model. A similar analysis which considered different assumptions for the variation of the cosmological constant was performed in \cite{2021PDU....3200807L}. Also in \cite{2021PhRvD.104l3512A}, the authors consider a cosmological model with sign switching cosmological constant and show that this model can alleviate some tensions that are related to the so called Hubble tension. In the present paper we perform  a  more   complete and strict  (methodologically  speaking)  analysis  of  the viability of the scenario studied in \cite{Perez:2020cwa}. Concretely  we  study a  relatively  simple model  parameterized by three quantities:   one characterizing  the  magnitude of the violation of energy conservation involved,  and the other two  caracterizing the  cosmological  period   during   which the  violation is assumed to have  taken place.
 We analyze the  model  by  simultaneously   considering, the anisotropies and polarization of the CMB, involving therefore the growth of the structure of the universe, in addition to the other  data related to  the  global geometry of the universe.

The physical  source of the   relatively  large violation  of the conservation of energy momentum  is thought to  be related to  the  dragging  effect  on the rotation of  black holes,   presumably arising from   some fundamental granularity of space-time.  Thus, in order to to extract  a  formula  for the
 {\it energy-momentum violation current} $\mathbf J_a \equiv \nabla^b  \mathbf T_{ab}$ and produce quantitative calculations, one would need to model  the cosmological  density of   black holes  as a function of their mass, their angular momentum and  cosmic  time  $n =  f (M, J, t)$   in combination  with  the proposed  form  of the effective dissipation rate described  in \cite{Perez:2019gyd} .  With that  expression at hand  we could  directly insert  it  into the  equations of UG, and study the resulting modifications of the late time  cosmology, allowing us to  check  whether or not the corresponding predictions can account well  for  all  the relevant observations.  
Unfortunately,  at this  stage, our  knowledge about  $f ( M, J, t)$  is  just too  vague  to  carry such     analysis.
There  are only some  known bounds on the proportion of  matter that might  be  in  the form of primordial black holes which unfortunately  seem to  refer mostly to  monochromatic  contributions \cite{2021RPPh...84k6902C}  rather  than constraints on  something like the function   
$f ( M, J, t)$ or even  the momentum integrated  function  $$F(M, t) =\int_0^{M^2} f(M, J, t) \ dJ $$ (the upper  bound  resulting  from limiting  consideration of up to the extremal  case).  Beyond that,  we have  recently learned \cite{2017EPJWC.16401011M,2021JPhG...48d3001G,2010PhRvD..81j4019C,2019NatAs...3..524N,1993Natur.366..242H,2018PhRvL.121n1101Z,2007ApJ...665.1277M,2014PhRvD..90h3514K,2013PhRvD..87l3524C}  that   the range of masses of black hole is   very wide,    involving in the lower end  objects as  light as    few solar masses  and at the other extreme  objects as large as $10^9 $ solar masses,    with   increasing   evidence for the existence of   many  objects   in the  60-100 solar mass range. 
 It is fair to say that we have  very little  understanding  regarding the process by which all but the  lighter  black holes formed  and also   very limited  knowledge of  the form that even  $ F(M,t) $  might take  at  intermediate  scales  (say $10^2-10^5$  solar mass range).
 
 Despite  our ignorance concerning the details of the current  characterizing the  violation  of  energy-momentum conservation  in the general  cosmological context, it is useful to perform a first analysis using recent cosmological data, such as the ones provided by the CMB, Baryon Acoustic Oscillations (BAO), SnIa and CC, to test the viability of cosmological models that are motivated by UG to explain the cosmological evolution including the growth of perturbations. However, it follows from the previous discussions that some assumptions,  related to the violation of the energy-momentum conservation,  will have to be  included in order to describe the behavior of  the dark energy component of the universe as well as the baryonic and dark matter ones.  

The organization of this manuscript is as follows:  In  section \ref{theory}  we offer  a brief discussion of the  general theoretical   ideas  underpinning this work.   Section  \ref{theory2}  is devoted to  specify the  concrete  cosmological model  based on these ideas  and  to characterize its  effective parametrization as  well as the ensuing   evolution of the   various contributions to the  energy density budget  as a function of cosmic time. We also discuss  in this section the effects of the modified cosmic evolution introduced by our model on the CMB spectrum.  The details of the analysis method used in this work to perform the statistical analyses (including the observational data sets)   are  discussed in section \ref{analysismethod}.   Results of the statistical analysis of the modified cosmological model motivated by unimodular gravity with CMB, BAO, SnIa and CC data are shown  in Section \ref{results}. We  end  in  section  \ref{conclusions}  with a   brief discussion of the results   the overall  viability   of  the model and  of the general theoretical scenario.

\section{Theoretical Scenario}.
\label{theory}

 A novel  and rather successful  scheme 
 to account for the  nature  and  magnitude of the cosmological  constant, yielding  in a natural  way its correct  order of   magnitude,  was proposed in \cite{Perez:2018wlo, SP2019}.  
 This was a  particularly encouraging result,  particularly  when contrasted  with  standard estimates   that are  typically   wrong by 120 orders  of magnitude. 
 The ingredients underlining  the proposal are:
\begin{enumerate}
 \item The generic  idea  that quantum gravity implied  some sort of spacetime  granularity.
 \item \label{tuti}The result of \cite{Collins-2004} where it was showed that the granularity cannot select a preferential reference frame without generating (presently) detectable violations to Lorentz invariance at low energies.
 \item The  idea  that the two  considerations  above might  be reconciled if the effects of granularity become relevant only in the presence of strong gravitational fields or high curvature.
  \end{enumerate}

 A    basic assumption  is that  the interactions with granularity need to be of a {\it relational } nature  in the  sense  that the relevant affected degrees of freedom must be capable (via their excitations) of  selecting the  local preferential  reference frame with respect to which the fundamental scale can have an operational meaning. 
This suggested the  rather  natural  assumption  that the probes which  could be  sensitive to such granularity needed to  posses  both,   a  scale (and thus break scale invariance to relate to the curvature scales) and an intrinsic  direction   to  determine the  directionality of the  effects under consideration. The natural candidate was thus identified as spinning and massive degrees of freedom. The effect would have to be modulated by curvature (to avoid the no-go result of item \ref{tuti} above). A mean-field perspective suggests taking the scalar curvature $ \mathbf R$ which (linked to the trace of the energy-momentum tensor via Einstein's equations) is a natural order  parameter of the violations of scale invariance for the gravitational sources involved. This led  us to propose that, at level  of  the  effective characterization    of matter in  terms of particles,    the effect  would  take  the form of a deviation from the  standard  geodesic  equation  given  by:
\be\label{modimodi}
u^{\mu}\nabla_{\mu}  u^{\nu}=\alpha \frac{m}{m^2_{\rm p}}\, {\rm sign}(s\cdot \xi) { \mathbf R}\, {s^{\nu}}, 
\ee
 where $\alpha> 0$ is a dimensionless constant,   $m$ is the  particle's mass, $m_P$ is Planck's mass, $s^\mu$ is the  particle's  spin, and $\xi^\mu$ is a preferred time-like vector field characterizing  the  state of motion  of the  matter   source.
The analysis  of this  hypothesis  in the cosmological context in terms of   standard kinetic theory led to  an effective violation of   energy momentum  conservation   given by \cite{Perez:2018wlo, SP2019}:

 \be\label{modimodi1}
{\mathbf J}_\nu  \equiv 8\pi G \nabla^{\mu}{\mathbf T}_{\mu\nu} =  4\pi \alpha  \frac{{T} }{m_p^2}R [ 8\pi G \Sigma_{i} |s_i| {\mathbf T} ^{(i)} ] \xi_\mu;
\ee
  where    $ {T} $ is the  temperature of the  cosmological plasma, the sum involves contributions of the different species $i$ in the standard model of particle physics,   ${\mathbf T}^{(i)}$ is trace of the energy momentum tensor of the   species $ i$, and   $ |s_i|$ the spin of that species.

   Moreover,  given  that  violations of the energy-momentum conservation are inconsistent in the context of GR, the proposal needs to be formulated in a context of a gravity theory that allows for such violations \cite{SP2019,PS2020}. In fact, the theory known as Unimodular Gravity permits a specific kind of violations and is therefore ideal to describe those  type  of effects expected to  arise from  a fundamental space-time  granularity. The relatively recent interest in this theory, which was initially considered by Einstein himself, can be traced back to works such as \cite{1989PhRvD..40.1048U,1989RvMP...61....1W}. At this  point  it  is  worthwhile  to   briefly  describe the 
theory of Unimodular Gravity. The action can  be  written   in a   coordinate  independent   form,  using, say,  abstract index  notation (see  for instance   \cite{Wald:1984}): 
 \begin{equation}
S = \int  [ R \epsilon_{abcd} + \lambda  (\epsilon_{abcd}  - \varepsilon^{}_{abcd}   )  +   {\cal L}_{\rm matt} \, \epsilon_{abcd}] 
\end{equation}
where $ \varepsilon^{}_{abcd} $   is a fiduciary  4-volume  element and  $ \epsilon_{abcd} $ is  the  4-volume  element  associated  to  the metric  
$ g_{ab} $, while $ \lambda (x) $   is a Lagrange multiplier function. Using coordinates  $x^{\mu}$ adapted to the  fiduciary  volume element one is led to the simple relation $\epsilon_{abcd} =\sqrt{-g} \varepsilon^{}_{abcd} $ with $g =  \det  g_{\mu\nu}$.
The  theory  is said to be  invariant under  a restricted  class  of  diffeomorphisms, i.e. those   that   are 4-volume  preserving\footnote{ Here, as is  common in these  discussions  it is  assumed that  only the dynamical elements  are   subject to the  action of the diffeomorphisms.}.
    The  resulting  equations of motion  for the metric are:
  \begin{equation} 
  G_{ab} (x) + \lambda  (x) g_{ab}(x)   =  8\pi G  T_{ab} (x) 
\end{equation}
while  the  variation with respect to $\lambda$ leads  to the constraint $ 
\epsilon_{abcd}  = \varepsilon^{}_{abcd}  $. The value of the Lagrange multiplier
 $\lambda(x)$  can   determined  by   taking the trace  of the previous equation. This leads to :
   \begin{equation}
\lambda(x)   =  \frac14  (8\pi G  T+ R  )
\end{equation}
 which  upon substitution result  in the   standard  form of the  equations of unimodular gravity,  namely: 
  \begin{equation} 
  R_{ab}   - \frac14  g_{ab} R  =  8\pi G  ( T_{ab} -\frac14  g_{ab}T)   
\end{equation}
 As noted a  central  feature of this  theory,  which makes it useful  for our purposes,  is that, in   contrast  with  general relativity,  it  does not  require the  conservation law $ \nabla^a T_{ab} =0$,  for its  self-consistency.
 In fact  one  might define the  `energy momentum non-conservation current' as  $J_a\equiv  8\pi G \nabla^bT_{ab}  \not=0$
 which  provided the integrability condition $dJ=0$  is  satisfied\footnote{The integrability condition is a direct consequence of the volume-preserving diffeomorphism invariance of the action principle \cite{Josset:2016vrq}.},   can be integrated   to yield,
 \begin{equation} 
{R_{ab} -  \frac 12 g_{ab} R   +   g_{ab}   
\left(   \lambda_{-\infty}  +  \int J   \right)
 =  8\pi G   {T_{ab}} }
\end{equation}
which coincides  with Einstein's equation  with a term  that   acts  like  an `effective dark energy component'  which need not be a constant. 
In the cosmological  setting,  the  role of   this  term---just as  it   occurs  with the   usual  cosmological constant term for the range of values 
 that are  relevant  in our universe---is entirely negligible  except for the very  late times,  where    every other  contribution to the  universe's   energy  budget  has  be  diluted to  an  extreme  level, and  thus  such  term  can   become    dominant.

 Calculating the contribution to dark energy in terms of  equation \eqref{modimodi1} using the standard the cosmological history of our universe \footnote{ This is  justified  because it can be    shown that  the resulting changes in the dynamics of the background are completely negligible during the period  in which the effective cosmological constant is  generated,  and  its  effects only become  relevant at the    very late    cosmological  times  when  the dilution of all other forms of  energies make the dark energy the dominant  component.},  it was found that the effective cosmological constant was given approximately by:
\be\label{modimodi2}
\Lambda \approx  16 \pi \alpha \sqrt { \frac{5\pi^3}{g^*}} \frac {m_t^4 T_{\rm EW}^3}{\hbar^2 m_p^2} \epsilon( T_{\rm EW})
\ee
 where $ m_t$ is the mass of the heaviest  particle with non-vanishing spin (here the top quark),  $T_{\rm EW}$ is the  temperature of the  electroweak transition,  $ g^*  $  is the degeneracy factor and  $\epsilon$ is  a factor that  takes into account the running of the top  quark mass. 
 For  standard values  of the cosmological parameters the previous prediction coincides  with the  observed value of the   cosmological dark energy in order of magnitude  when the dimensionless parameter $\alpha$ is assumed to be of  order  $1$.  
 
 It  became  quite clear  after subsequent  analysis that, apart from the generation of a cosmological constant, the  conditions for other effects  to  be  observable   where simply to extreme to be accesible to testing in situations to which we have access at present in our universe. For instance, simple dimensional considerations imply violation of Lorentz invariance parametrized in dangerous\footnote{These  are  terms that  might appear in the effective  matter Lagrangian,  which  have the potential to   generate violations of  Lorentz  Invariance  that  are  large enough  to  enter  in conflict  with   empirical bounds. } dimension 5 operators modulated by  $\mathbf R$  such as:
\be
\lambda \bar \psi \gamma_\mu \psi \xi^\mu \frac{{\mathbf R}}{m_p},
\ee
for fermion fields  $\psi$,  a dimensionless $\lambda$.
According to Kostelecky \cite{Kostelecky:2008ts} the strongest bound on such operators is
\be
\frac{\lambda {\mathbf R}}{m^2_p}\approx \lambda \frac{\rho}{m_p^4}<10^{-12}.
\ee
With neutron star densities estimated to reach scales of  about $\rho_{\rm n}\approx 10^{-80} m_p^4$, it is  easy to see that not even   under such  extreme  conditions,  empirically   significant  direct manifestations of the previous fundamental matter-diffusion-type of effect can take place in  matter configurations as presently known. 
 
However, similar effects could be important in the context of non-fundamental matter forms. For instance in the case of black holes produced by the gravitational collapse of fundamental matter fields which latter become (for the purpose of outside observers) seemingly empty region but with a (classically) unbounded internal curvature. Of  course,  black holes  are not  simple point-like particles   and  the applicability of  equation \eqref{modimodi1}  to such macroscopic objects might not seem directly justified.  Nevertheless, the fact that black holes are intimately connected  with regimes  of arbitrarily high curvatures---expected  reach  Planck scale curvatures inside where only a  full quantum gravity description could be  truly  relied on---naturally opens the door to the consideration of the phenomenological possibility  that they might be  affected in a special  way  by  our hypothetical spacetime  granularity.  
 This  suggests, as  discussed in \cite{Perez:2019gyd} that  black holes  could be subject  to  some   kind of {\it effective  friction}  related to the motion of the ``frames  they  drag"  with   them  and  those  associated to the matter that was connected to the  space-time structure of their respective environment.
 
 This led  to  two   interconnected effects responsible for  a  translational and  a rotational friction respectively.  An early analysis  carried out in  \cite{Perez:2019gyd}  showed  clearly that the  one  that seems  more likely to have  any  important   energetic  relevance  was  the second  effect,  in which (assuming,   conservatively, that the  second law would not be  violated by the type of effects considered)  up to 30\% of the mass of  a   highly  rotating  black hole  could be lost  as  a result of  the  rotational friction.   Moreover, the same analysis indicates that in the  context of unimodular gravity the rotational friction  could be  a significant source of a  change in the effective cosmological constant.

 Even when the basic theoretic ingredients for the generalization of \eqref{modimodi1} to the case of black holes are available from the analysis of \cite{Perez:2019gyd},  an accurate account of the number  density of  black holes  as a function of  their mass  angular momentum and  cosmic  time  $ n= f (M, J, t)$  for quantitative estimates. Such formula presumably would have to incorporate the formation of new black  holes  as the  result of  standard gravitational collapse during cosmic evolutions, the effects of black hole mergers,  as well as  any possible contribution of primordial black holes.  With  $n= f (M, J, t)$ one would be able to   compute, both, the modification of that distribution due to the  novel  friction hypothesis, as  well as (most importantly for this work) its  exact  contribution to the evolution of the dark energy density during late cosmological dynamics.  A  project is underway to   try  to construct  reasonable models  for   $f (M, J, t)$ taking into account hypothetical  primordial distribution,   simplified  models of merger  histories, and  direct  bounds on abundances  of black holes  at   certain mass  scales. 
    
Unfortunately  that task  will not   be  completed  in the  very   near future to the   level  where the form of $ f$ will be narrow  enough   so that   we   could   think  of simulating  the effects  with a single  $  f ( M,  J, t)$. That means that  even in the best of  circumstances we  will have a  multi-parametric  characterization of the possible  $f ( M,  J, t)$'s  and the  studies  of their cosmological effects  will have to be   performed  in conjunction with data analysis  of many other sources  such as the one   we  will contemplate here.

 In view of this, one can consider rather simple models to describe the evolution  of the cosmological constant over time \cite{Perez:2020cwa}.  As this  study is motivated  by the so called  ``$ H_0 $ -tension" we focus attention  on the  result of the effect we have described  that  might  have  taken place from the time of decoupling until today. We therefore study  the overall feasibility  of the model in the light of the available cosmological data.

Some preliminary results were obtained by setting the current value of Hubble parameter,$H_0$, to the SnIa estimation \cite{2019ApJ...876...85R}
and considering  for the range of values of the model parameters  that  would  be roughly compatible with the observational value of $\theta$, the angular acoustic scale of the CMB. This gave encouraging results  in the light of the  $H_0$ tension, 
as the  fitting of the  models parameters  did not require a  large  change  in either, the   age of the universe, or the present amount of matter (both baryonic and dark components). 
We note that this preliminary analysis only studied  global geometric quantities  like $\theta$, however a change in the evolution of  cosmic history introduced by this kind of models would  also affects the growth of perturbations that are the seeds of the present cosmic structure.  In order to  include  consideration of such aspects, a more complete analysis, like the one presented in the present paper, that  also includes the prediction of the anisotropies and polarization of the CMB, {as well as geometric quantities such as the luminosity or angular diameter distance,  is necessary.

In this work, we consider a  some relatively  richer model which will be carefully described in section \ref{theory2}. The friction-like effect  will be characterized by three parameters: two of them are used to describe the period in which the  effect of  `dissipation of black hole rotational energy'  takes place, and the last one characterizes the   overall magnitude of the corresponding  change in the  cosmological constant.   At the level of analysis  that is  possible,  given  our lack of  knowledge regarding  $ f(M,J, t)$,  it makes sense to  simplify   the treatment and  model  directly the  time dependence of   the  effective energy transfer to dark energy  to study its  effects in the  dynamics of  the  scale factor.
These modifications of the background evolution are then inserted  into the  standard analysis  of    structure  growth   starting from  the  usual form of the   quasi flat   scale  invariant primordial spectrum usually taken to    emerge  during the inflationary epoch.

This paper aims at dealing with two crucial aspects. First, the constraining of the parameters of the theory with the current data using a complete and  appropriate statistical analysis. Next, the  examination and interpretation of the results obtained in light of other  relevant  constraints.  In particular  one has 
the  usual  nucleo-synthesis constraint  setting the  overall contribution to  the      universe's  current energy budget   appearing in the form of   baryonic  matter  at  $4$-$5 \%$, and on the other    the high  precision data  emerging from  the   CMB, providing bounds on  the  relative abundances  of   dark  matter to baryonic  matter  at the last scattering surface. 
On the other hand,  as the model  calls for  anomalous  reduction of the corresponding   energy densities at late times, we must ensure compatibility with  the data from nearby  galaxies and  galaxy clusters constraining the present abundance of baryonic matter as well as   bounds on non luminous  mater (that  might include  some  baryonic components). In other words, the proposed mechanism, if it is to  be  considered as viable,   should not result in  too large of a reduction in baryonic  or  dark  matter components  today  so as  to  generate  a conflict  with the relevant observations. As we  will see  the best fits  obtained in this  work  avoid that {\it `danger'} by a  rather large margin.

\section{The matter densities in the unimodular model}
\label{theory2}

 Let  us  consider  a   homogeneous  and isotropic Friedmann-Lema\^{\i}tre-Robertson-Walker (FLRW)  universe, 
 and introduce our modifications at an effective level in the sector of both matter (including baryonic  and  dark matter) and dark energy. In the context of the UG theory, the modified conservation equation can be expressed as:
%
\be
\dot \rho_M  +  3 \frac{\dot a}a \rho_M   = -\dot \rho_{\Lambda}
\label{eq:rhomatter}
\ee  
Therefore, provided an expression for $\rho_\Lambda (t) $ we can integrate the latter equation from the beginning of the radiation era ($t_{\rm rad}$)  up   to    an arbitrary   later time,  $ t$, to  obtain an expression for $\rho_M (t)$. As discussed previously,  even  with a  phenomenological  model for the diffusive process in black holes  such as that described in the  Appendix \ref{BH_effects} (and  originally in\cite{Perez:2019gyd}), the form  of the energy-momentum violation current cannot be well described given our present  lack of  knowledge of the black holes densities in the late universe. However, we can  postulate a simple expression for $\rho_\Lambda $ in order  to analyze the generic  viability of a UG model with difussion as follows 

  \be\label{DE}
 \rho_{\Lambda} (a)  =  \rho_{\Lambda}(t_{\rm rad})   + f(a) \Delta\rho
\ee 

At $t_{\rm rad}$, the corresponding scale factor is referred as $a_{\rm rad}$. Assuming  $f (a_{\rm rad} )=0$  and  $f (a_0)=1$,     the current (i.e today) value of the dark energy  density  reads: 
\be
\rho_{\Lambda} (a_0)  =  \rho_{\Lambda}(t_{\rm rad})   +  \Delta\rho= \rho_{\Lambda}^0
\ee
where $\rho_\Lambda^0$ is the current value of the dark energy density. The final expression for $\rho_{\Lambda} (a)$ yields
\be\label{DE_2}
 \rho_{\Lambda} (a)  =  \rho_{\Lambda}^0   +  \Delta\rho \left[f(a) - 1 \right]
\ee

  For the sake of simplicity we assume $ f(a)$  to  be  constant  in two time intervals, while shows a linear behaviour in the interval of interest as follows:
\begin{equation}
	f(a) = 
						\begin{cases}
							\quad 0 \quad\quad \quad \quad \quad \quad \quad \quad \quad  a \, \in \,\, (a_{\rm rad},a^*-\delta/2) \,, \\ \\
							\quad  \dfrac{a- a^* +\delta/2}{\delta} \quad \quad \quad  a \,\in \,\,  (a^* -\delta/2, a^* +\delta/2) \, \\ \\
\quad 1 \quad\quad \quad \quad \quad \quad \quad \quad  \quad  a \, \in \,\, (a^*+\delta/2,a_0) \,, 
						\end{cases}
\label{fa}
\end{equation}

In this way $\rho_{\Lambda}$ is constant during the universe evolution, except for the time interval  for which the value of the scale factor $a$ is $(a^* -\delta/2, a^* +\delta/2)$ where it changes linearly in $a$.  Moreover, the value of the cosmological constant is not always the same, namely $\rho_{\Lambda}= \rho_{\Lambda}^0   -  \Delta\rho$ when $a  \in \,\, (a_{\rm rad},a^*-\delta/2)$ and $\rho_{\Lambda}= \rho_{\Lambda}^0$ when  $a  \in \,\, (a^*+\delta/2,a_0)$.
We  should   consider  of course $   a_{\rm rad} <  a^* -\delta/2 $ and   $a^*  + \delta/2 < a_0$. In this way, $\Delta \rho$ accounts for the change in the value of the cosmological constant with respect to its present value $\rho_{\Lambda}^0$, and $(a^*-\delta, a^* + \delta)$ refers to the time interval in which the energy-momentum tensor is not conserved. 
\medskip
Consequently, we can obtain the  expression for  the  total matter density $\rho_M(t)$ as a function of the scale factor integrating Eq. \ref{eq:rhomatter} and assuming Eqs. \ref{DE_2} and \ref{fa}:

\begin{equation}\label{totalmatterdensity}
	\rho_M (t)\quad = \quad
						\begin{cases}
							\quad  \dfrac{a^3_{\rm rad} \, \, \rho_M (t_{\rm rad})}{a(t)^3}   \quad\quad \quad \quad \quad\quad \quad \quad \quad \quad  \quad\quad   a \quad \in \,\, (a_{\rm rad},a^*-\delta/2) \,, \\ \\
\quad    \dfrac{a^3_{\rm rad} \, \, \rho_M (t_{\rm rad})}{a(t)^3} -  \dfrac{\Delta\rho}{ 4 \delta} [ a  -  \dfrac{(a^*-\delta/2 )^4}{a(t)^3}  ]      \quad \quad  a \quad \in \,\,\,\, (a^* -\delta/2, a^* +\delta/2) \,,  \\ \\
							\quad      \dfrac{a^3_{\rm rad} \, \, \rho_M (t_{\rm rad})}{a(t)^3}  - \dfrac{\Delta\rho}{a(t)^3} [ {a^*}^3  +  \dfrac{(a^*\delta^2)}{4}]           \quad \quad a \quad \in \,\, (a^*+\delta/2,a_0)  \quad \,. 

						\end{cases}
\end{equation}

We now proceed to analyse the observational predictions of this model and how the CMB spectra are modified with respect to the standard model as the $a^*$, $\delta$, $\Delta {\rho}$ parameters vary.
For this, we recall  that  baryons and dark matter have different physical interactions with other particles  during the formation of neutral hydrogen and  the  following universe evolution, and  therefore  in principle they should be  described separately}.
Let us propose the following expression for the baryon, $\rho_B (t)$, and dark matter, $\rho_{DM}(t)$, densities which follow from Eq. \ref{totalmatterdensity}

\begin{eqnarray}
\rho_B (a) &=&  \frac{a^3_{\rm rad} \, \, \rho_B (t_{\rm rad})}{a(t)^3} + \alpha F(a,\Delta \rho, a^*,\delta) \label{baryonmatterdensity}\\
\rho_{DM} (a) &=&  \frac{a^3_{\rm rad} \, \, \rho_{DM} (t_{\rm rad})}{a(t)^3} + (1 -  \alpha) F(a,\Delta \rho, a^*,\delta) \label{darkmatterdensity}
\end{eqnarray}
where
\begin{equation}\label{Fa}
	F(a,\Delta \rho, a^*,\delta) = 
						\begin{cases}
							\quad  0  \quad\quad \quad \quad \quad\quad \quad \quad \quad \quad   a \quad \in \,\, (a_{\rm rad},a^*-\delta/2) \,, \\ \\
\quad    -  \dfrac{\Delta\rho}{ 4 \delta} [ a  -  \dfrac{(a^*-\delta/2 )^4}{a(t)^3}  ]  \quad  \quad       a \quad \in \,\,\,\, (a^* -\delta/2, a^* +\delta/2) \,,  \\ \\
							\quad       - \dfrac{\Delta\rho}{a(t)^3} [ {a^*}^3  +  \dfrac{(a^*\delta^2)}{4}]  \quad        \quad   a \quad \in \,\, (a^*+\delta/2,a_0)  \quad \,. 
						\end{cases}
\end{equation}   
In order to obtain expressions for $\alpha$, $a_{\rm rad}, \, \, \rho_B (t_{\rm rad}),\,\, \rho_{DM} (t_{\rm rad})$ in terms of the present values of the baryon density parameter $\Omega_B = \frac{\rho_B(t_0)}{\rho_{\rm crit}}$ and the dark matter density parameter $\Omega_{DM} = \frac{\rho_{DM}(t_0)}{\rho_{\rm crit}}$ where $\rho_{\rm crit}$ is the present critical density, we first  define  the present fraction of   baryon to dark matter density as follows

\begin{equation}
\beta = \frac{\rho_B(t_0)}{\rho_{DM}(t_0)}
\label{beta}
\end{equation}

Next, we assume that the fraction of baryon to dark matter at the beginning of the radiation era is equal to its present value\footnote{As  a  rather straight forward  motivation for this  simplifying assumption  we might adopt the view that  most black holes in the universe   ( or  at least those that play  a substantial role in   the process  at hand) originate ultimately in a population of primordial black holes,  which in turn where formed well  before  the  spatial distributions of  dark and  baryonic matter  differed  substantially.  Thus   the great majority of  matter in the form of black holes can be viewed  as traceable  to   such  components which  would  therefore naturally  contribute   in proportion to their cosmic  abundances.   The result is that, when    energy  in   black holes is  lost  as a result of  our hypothetical  aforementioned  ``effective  friction”, the    proportion of baryonic and  dark matter would   remain essentially unchanged   in the whole process.}  
\begin{equation}
\beta = \frac{\rho_B(t_0)}{\rho_{DM}(t_0)}=\frac{\rho_B(t_{\rm rad})}{\rho_{DM}(t_{\rm rad})}
\label{beta_2}
\end{equation} 
{With a bit of algebra we obtain from Eqs. \ref{beta} and \ref{beta_2} that}

\begin{equation}
\alpha= \frac{\beta}{1 + \beta} = \frac{\rho_B(t_0)}{\rho_B(t_0) + \rho_{DM}(t_0)} = \frac{\Omega_B}{\Omega_B + \Omega_{DM}}
\end{equation}

It is important to note that $\Omega_B$ and $\Omega_{DM}$ are different from $\Omega_B^{\Lambda}$ and  $\Omega_{DM}^{\Lambda}$ the current baryonic and dark matter densities in units of the critical density defined in the standard cosmological model.

Therefore, recalling that $\Omega_B = \dfrac{\rho_B(a_0)}{\rho_{\rm crit}}$ and  $\Omega_{DM} = \dfrac{\rho_{DM}(a_0)}{\rho_{\rm crit}}$ we obtain the expressions for $\rho_B(a)$ and $\rho_{DM}(a)$ in terms of $\Omega_B$ and $\Omega_{DM}$

\begin{equation}
\rho_B (a)= 
\begin{cases}
\dfrac{\Omega_B \rho_{\rm crit} + \dfrac{\Omega_B}{\Omega_B + \Omega_{DM}} \Delta \rho \, ({a^*}^3  +  \dfrac{a^*\delta^2}{4})}{a^3} \hspace{5.5cm} a \quad \in \,\, (a_{\rm rad},a^*-\delta/2) \, \\ \\
 \dfrac{\Omega_B \rho_{\rm crit} + \dfrac{\Omega_B}{\Omega_B + \Omega_{DM}} \Delta \rho \, ({a^*}^3  +  \dfrac{a^*\delta^2}{4})}{a^3}- \dfrac{\Omega_B}{\Omega_B + \Omega_{DM}} \dfrac{\Delta\rho}{ 4 \delta} [ a  -  \dfrac{(a^*-\delta/2 )^4}{a^3}  ]  \quad  \quad     \quad \quad  a \quad \in \,\,\,\, (a^* -\delta/2, a^* +\delta/2) \,,  \\ \\
\dfrac{\Omega_B \rho_{\rm crit}}{a^3} \hspace{9cm} a \quad \in \,\, (a^*+\delta/2,a_0)  \quad \,
\end{cases}
\label{omegab}
\end{equation}
\begin{equation}
\rho_{DM}(a) = 
\begin{cases}
\dfrac{\Omega_{DM} \rho_{\rm crit} + \dfrac{\Omega_{DM}}{\Omega_B + \Omega_{DM}} \Delta \rho \, ({a^*}^3  +  \dfrac{a^*\delta^2}{4})}{a^3} \hspace{5.5cm} a \quad \in \,\, (a_{\rm rad},a^*-\delta/2) \, \\ \\
\dfrac{\Omega_{DM} \rho_{\rm crit} + \dfrac{\Omega_{DM}}{\Omega_B + \Omega_{DM}} \Delta \rho \, ({a^*}^3  +  \dfrac{a^*\delta^2}{4})}{a^3}- \dfrac{\Omega_{DM}}{\Omega_B + \Omega_{DM}}  \dfrac{\Delta\rho}{ 4 \delta} [ a  -  \dfrac{(a^*-\delta/2 )^4}{a^3}  ]  \quad  \quad     \quad \quad  a \quad \in \, (a^* -\delta/2, a^* +\delta/2) \,,  \\ \\
\dfrac{\Omega_{DM} \rho_{\rm crit}}{a^3} \hspace{9cm} a \quad \in \,\, (a^*+\delta/2,a_0)  \quad \,
\end{cases}
\label{omegadm}
\end{equation}

On the other hand, we point out that the baryon and dark matter fractions that must be considered for the nucleosynthesis predictions are

\begin{eqnarray}
\Omega_B^{\rm early}&=& \Omega_B  + \dfrac{\Omega_B}{\Omega_B + \Omega_{DM}}\frac{\Delta \rho}{\rho_{\rm crit}} ({a^*}^3  +  \dfrac{a^*\delta^2}{4}) \\
\Omega_{DM}^{\rm early}&=& \Omega_{DM}  + \dfrac{\Omega_{DM}}{\Omega_B + \Omega_{DM}} \frac{\Delta \rho}{\rho_{\rm crit}} ({a^*}^3  +  \dfrac{a^*\delta^2}{4})
\end{eqnarray}

Note that,  except for the interval $(a^* - \frac{\delta}{2},a^*+\frac{\delta}{2})$, the baryon and dark matter densities behave in the same way as  the ones of the $\Lambda {\rm CDM}$ model.  Moreover, when   $a<a^*-\frac{\delta}{2}$, the  density parameters are equal to $\Omega_B^{\rm early}$ for baryons and $\Omega_{DM}^{\rm early}$ for dark matter, while for $a>a^* + \frac{\delta}{2}$,  $\Omega_B$ and $\Omega_{DM}$ are the ones defined above, and
  correspond to the ``late  time values". On the other hand, the evolution of the baryonic and dark matter densities during the interval  $(a^* - \frac{\delta}{2},a^*+\frac{\delta}{2})$ is different from the usual $\Lambda {\rm CDM}$ ones as can be inferred from Eqs. \ref{omegab} and \ref{omegadm}, which reflect what should be expected for the time interval during which the energy-momentum conservation is violated.

Let us stress, that we are considering the simplifying  assumption that both components of matter (baryonic and dark) are at the end  equally affected     by the    energy diffusion under consideration. The expressions we have obtained  for the evolution of the baryon and dark matter and dark energy densities in our unimodular model with diffusion will result in a modification of the  the Friedmann equation as follows:

\be\label{Friedman2}
H^2 (a)  = \frac{ 8\pi G}{3}[  \rho_{R}(a)  + \rho_B(a) + \rho_{DM}(a)  + \rho_{\Lambda}(a)] -   \frac{k}{a^2}
\ee

where $\rho_B(a)$, $\rho_{DM}(a)$ and $\rho_{\Lambda}(a)$ are described by  Eqs. \ref{omegab}, \ref{omegadm} and \ref{DE}  respectively and $\rho_{R}(a)$ is the radiation density which is not modified in our model. The change in the Hubble factor results in changes in other relevant physical quantities  for CMB physics such as 
$z_{\rm eq}$, i.e the redshift at which the density of matter and radiation are equal, $z_{LS}$, that is the redshift of last scattering,  $\theta(z_{LS})$, that is the angular diameter distance at last scattering and  $l_D$,  the diffusion damping length. Also, for baryon acoustic oscillations (BAO) physics we have to consider the modification in $z_{\rm drag}$, i.e the redshift at which baryons decouple from photons \footnote{Since there are far more photons than baryons,
after photon decoupling the photons continued to drag
baryons with them slightly longer into the Compton drag
epoch. The
redshift at which the baryon velocity decouples from the
photons is called $z_{\rm drag}$.}.  In turn, $z_{\rm drag}$  and $l_D$ are not only affected by the modification in the Hubble factor, but also depend on the speed of sound, which in turn depends on $\rho_B$.
However, we have checked that the variations of the latter quantities are of order $0.07$ \% for $\frac{\Delta \rho \, h^2}{\rho_{\rm crit}} =0.002$ and therefore the most important effects introduced by our model is the change in the Friedmann equation.

\subsection{Considerations for the CMB anisotropy and polarization spectra}
\label{Cl}
We implement the model discussed in the previous section in the public Code for Anisotropies in the Microwave Background (CAMB) \cite{Lewis:1999bs}, changing the expressions of $ \rho_B$, $\rho_{DM}$ and $\rho_{\Lambda}$ to the ones described in Eqs. \ref{DE}, \ref{fa}, \ref{omegab} and \ref{omegadm}, both in the background evolution and in the calculation of the growth of perturbations. Next, in order to compare the behaviour of the class of models analyzed in this paper,  we assume as a reference a fiducial model, namely a  $\Lambda$CDM one, with the cosmological parameters fixed to the bestfit values of Planck Collaboration (2018) \cite{planckcosmoparams18}.  Also, we define 

\begin{equation}
    \Delta \rho_{\Lambda}=\frac{8 \pi G}{3} \frac{\Delta \rho}{100^2} =\frac{\Delta \rho \, h^2}{\rho_{\rm crit}}
\end{equation}
that we use instead of $\Delta \rho$ to analyze the effects of assuming the unimodular models in the CMB anisotropy and polarization spectrum 
\footnote{In this way,the difference between the values of the cosmological constant at early and late times can be expressed as a density parameter, in a similar way, to the 
baryon and dark matter density parameters}.

\begin{figure}
\centering
\includegraphics[width=1\textwidth]{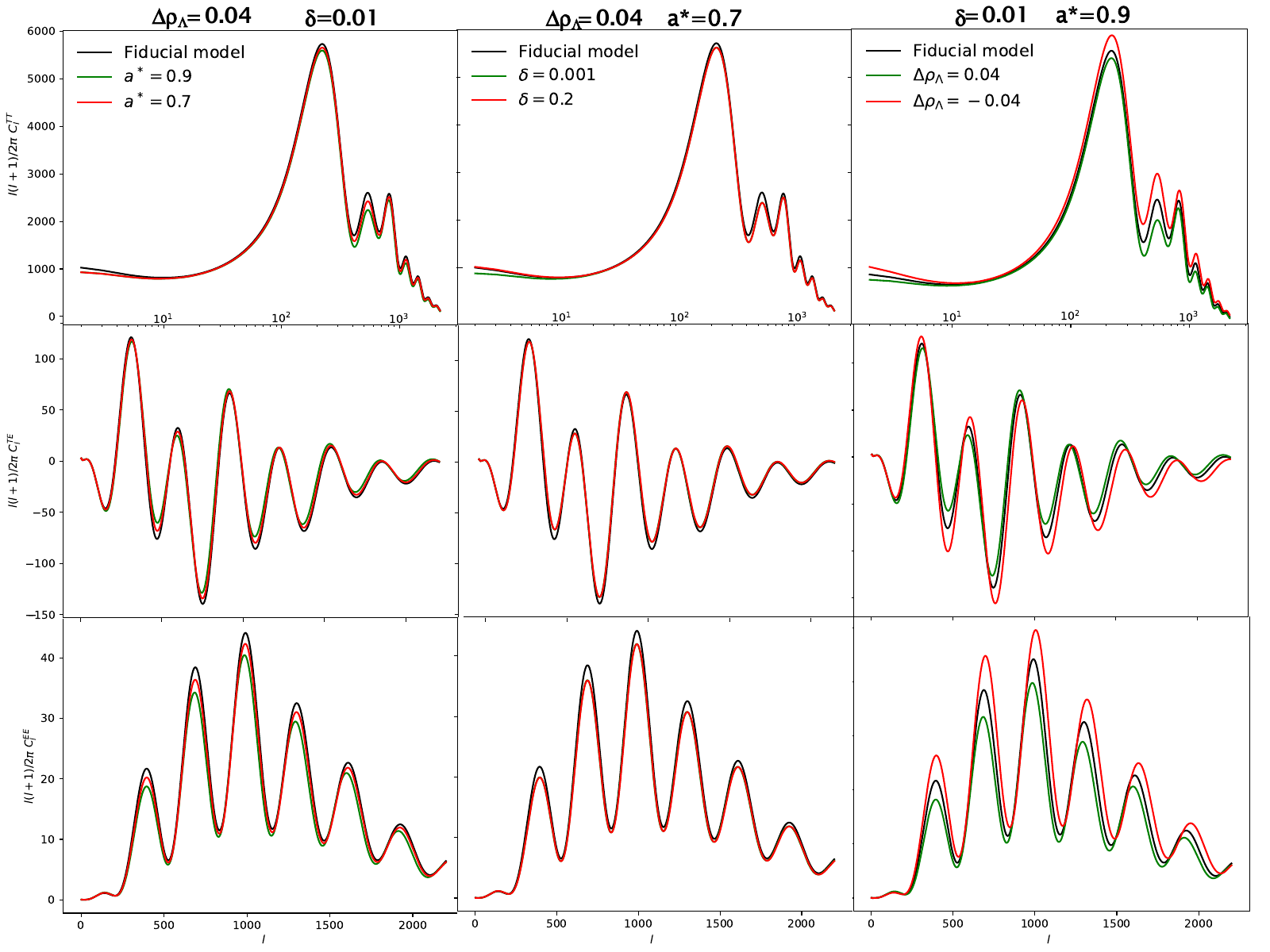}
\caption{The CMB anisotropy and polarization spectrum for different combination of the unimodular parameter. The two fixed parameters are indicated at the top of the column, while the legend of each column indicates the third parameter with two explored values. Top: Temperature auto-correlation function ($C_l^{TT}$); Middle: Temperature E-Mode cross-correlation function ($C_l^{TE}$); Bottom: E-model auto-correlation function ($C_l^{EE}$). }
\label{Clastar}
\end{figure}

In Fig. \ref{Clastar} we show the unimodular behaviour as the parameters $a^*$, $\delta$, $\Delta {\rho}_{\Lambda}$ vary, exploring the impact of one parameter at a time and leaving the other two fixed (indicated at the top of each column), while for the cosmological parameters we set the $\Lambda$CDM best-fit model reported by the Planck Collaboration (2018) \cite{planckcosmoparams18}. Our choice of the cosmological parameters for the fiducial and unimodular model results in that the physics of the late universe  ($a>a^* + \delta/2$) is the same in both models (see Eqs. \ref{omegab},   \ref{omegadm}, \ref{DE} and \ref{fa}) \footnote{Here we mean that both the behaviour of the energy densities and the corresponding energy parameters are the same in this period($a>a^* + \delta/2$). Note that as a result the unimodular model will not necessary fit BBN constraint, but the point of this section is to study the behaviour of the CMB spectra as the unimodular parameters are changed in order to understand the results of the statistical analysis of the next section. }.

We note that increasing $\Delta \rho_{\Lambda}$ results in a 
 decrease in the height of the Doppler peaks and a corresponding increase in the valleys. Also, we note that this produces   in addition, a very  small shift in the  locations of the  peaks. These effects are very similar to the ones that occur if the value of $\Omega_{DM} h^2$ is changed, so  we can anticipate a degeneration between $\Omega_{DM} h^2$  and $\Delta \rho_\Lambda$, in the statistical analysis and parameter constraints. Morevoer, it is useful to recall  that a decrease in the value of $\Omega_B$ has the effect of decreasing the height of the  odd peaks and enhancing the height of the even peaks. The latter is relevant, because we also note that  the decrease (due to the increase in $\Delta \rho_{\Lambda}$) in the Doppler peakes  is larger in the even  than in the odd  ones.  The reason for this, is that modifying $\Delta \rho$ affects both $\rho_{DM}(t)$ and $\rho_B(t)$ in a proportional  way  and therefore the effect that is shown in Fig. \ref{Clastar} is a combination of the change in both densities through the variation in $\Delta \rho$. Therefore, we also expect a degeneration between $\Omega_B$ and $\Delta \rho_\Lambda$, in the statistical analysis. 
 We further note that, as $a^*$ moves away from 1 in the  model, the stage of the universe during which its physical parameters are different from the fiducial model is enlarged, resulting in a decrease in the peaks and valleys of the spectrum. Moreover, a change in $\delta$ affects only  the lower multipoles of the CMB spectra which is usually  attributed to the Integrated Sachs-Wolff (ISW) effect if a standard model of inflation is assumed. Moreover, it is well known that the ISW effect scales with the amount of dark energy \cite{2014PTEP.2014fB110N}. We recall that the dark energy in our model behaves as $\frac{1}{\delta}$ (see Eqs. \ref{DE_2} and \ref{fa}) and this explains why for greater values of $\delta$, the departure from the standard behaviour is less than for lower values. Likewise, the low multipoles are also modified by a change in $\Delta \rho_{\Lambda}$ and $a^*$, and this can be explained since an change in these parameters also affects the amount of dark energy.
Finally, due to the lower sensitivity of the spectra to changes in $a^*$ and $\delta$, we expect the degeneration of these parameters with the  usual  cosmological ones to be small or almost negligible. 


\begin{figure}
\centering
\includegraphics[width=0.7\textwidth]{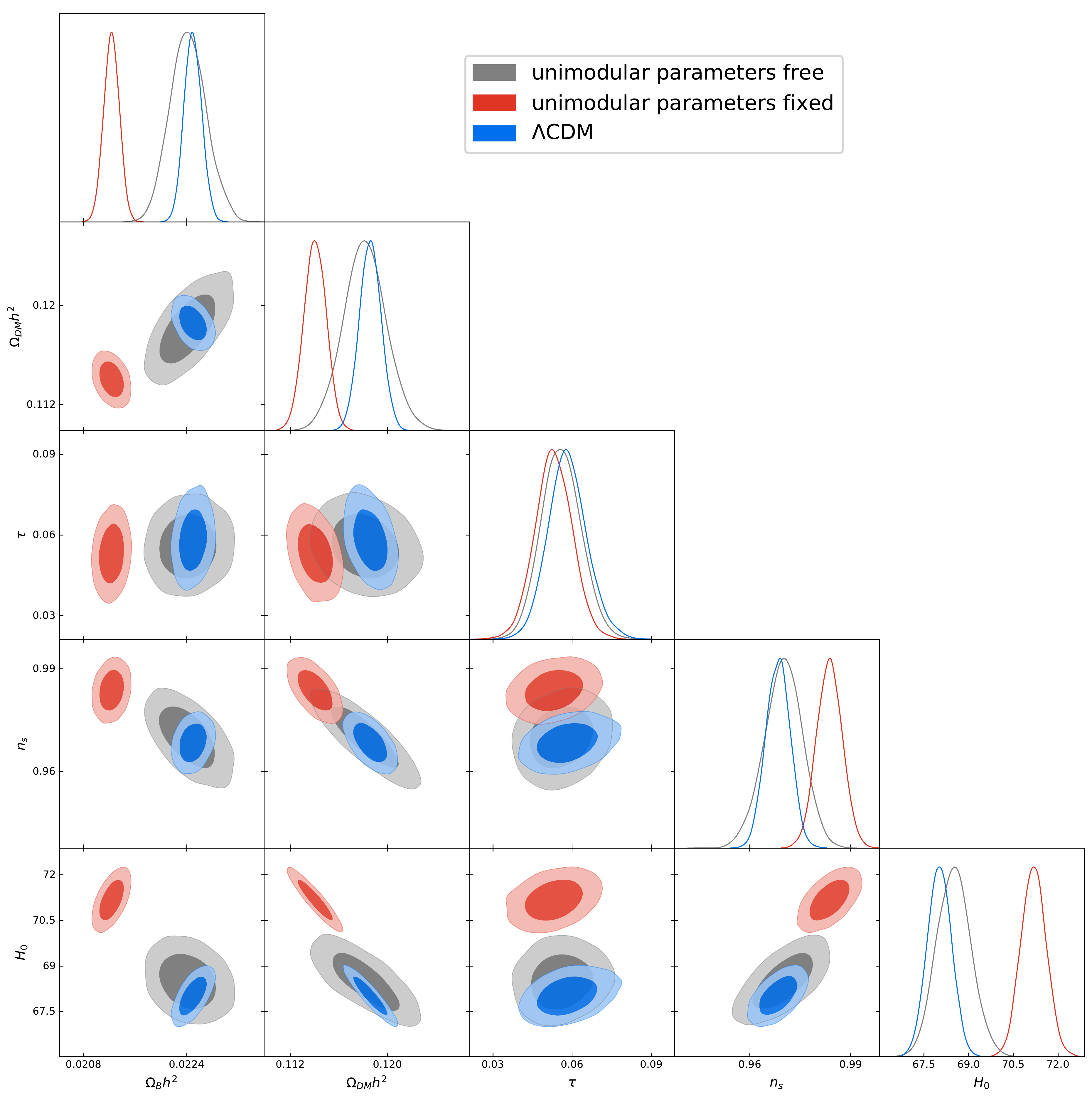}  
\caption{Constraints on model cosmological parameters considering the case of all parameters free to vary (gray line), setting theory unimodular parameters to arbitrary values (red line), compared with the standard cosmological model (blue line).}
\label{fig:tri_cosmo}
\end{figure}

\begin{figure}
\centering
\includegraphics[width=0.7\textwidth]{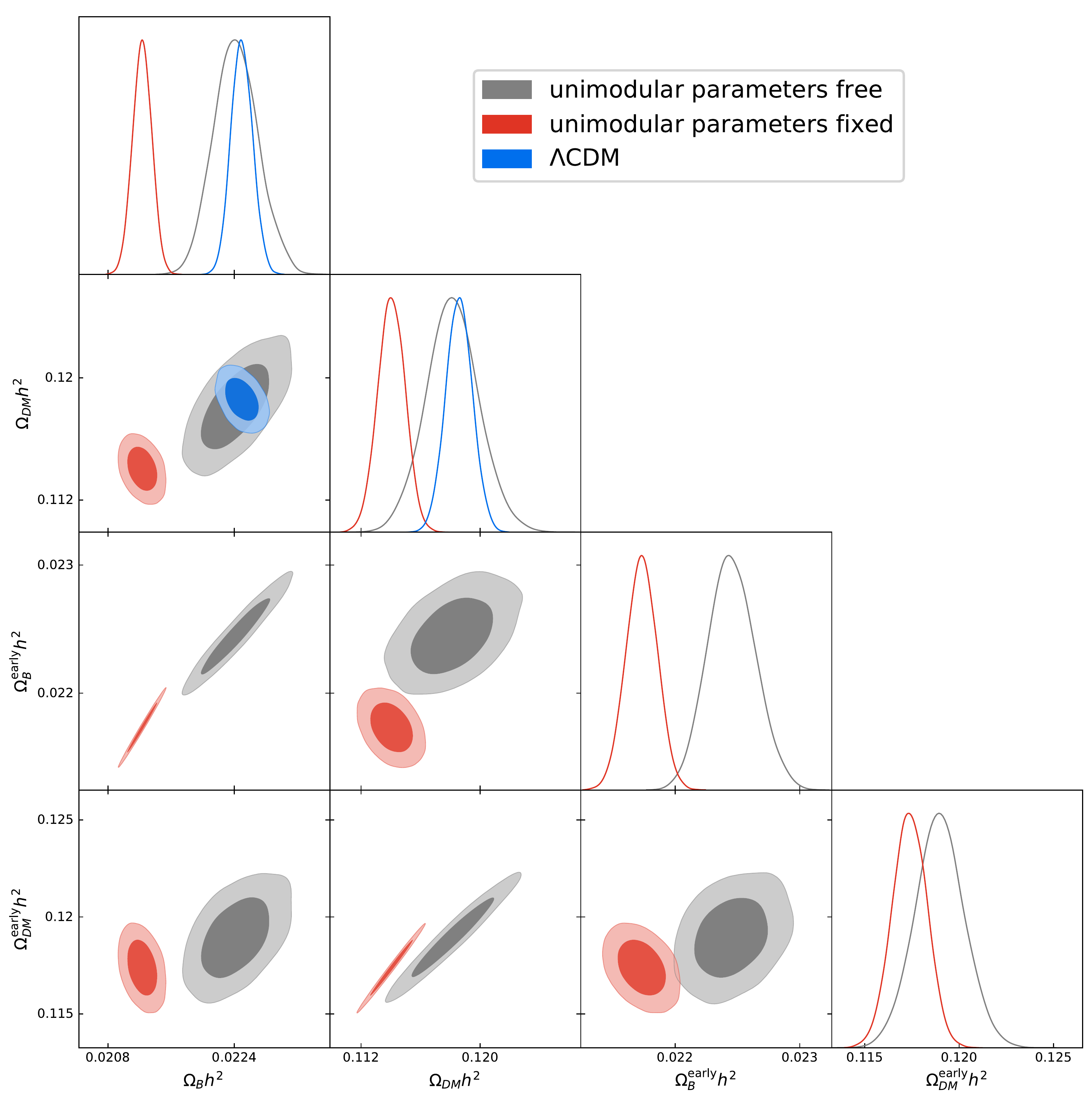}  
\caption{Constraints on model densities parameters ($\Omega_i h^2$) considering the case of cosmological and unimodular theory parameters free to vary (gray line), setting theory unimodular parameters to arbitrary  values ($\Delta \rho_\Lambda = 0.09$, $a^*=0.04$, $\delta=0.02$) while the cosmoslogical parameters are free to vary (red line), compared with the standard cosmological model (blue line).}
\label{fig:tri_omegas}
\end{figure}

\begin{figure}
\centering
\includegraphics[width=0.45\textwidth]{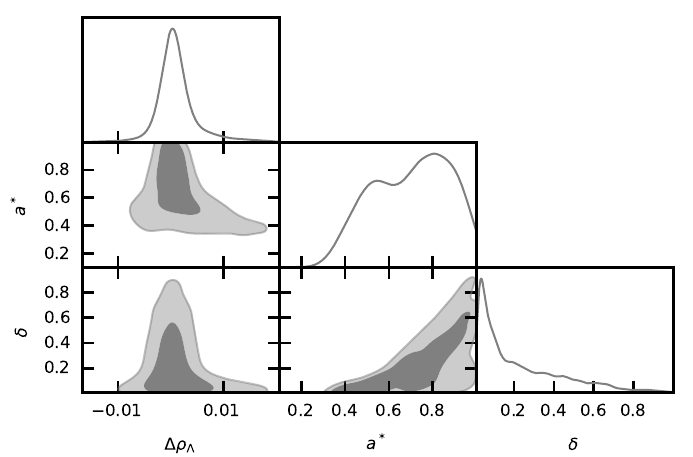}  
\includegraphics[width=0.45\textwidth]{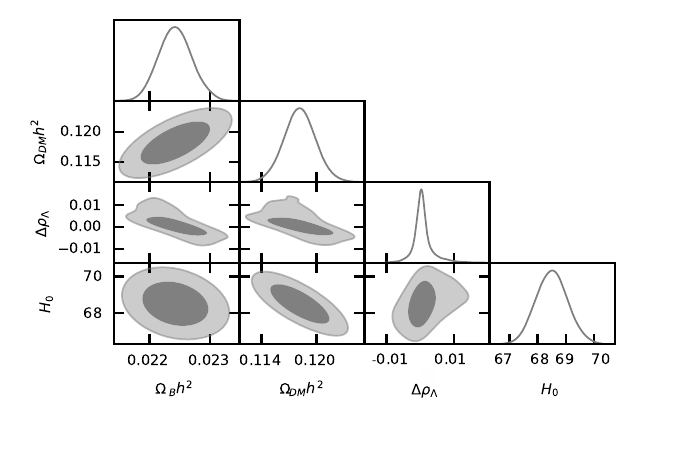} 
\caption{Constraints on model unimodular parameters.}
\label{fig:tri_unimodular}
\end{figure}

\section{ Methodology of the Analysis}
\label{analysismethod}
The observational predictions of the previous section are now compared with cosmological data, so as to obtain the constraints of the free parameters  of the unimodular model. To do this, we use the a Monte Carlo
Markov chain exploration of the parameters space using the available package CosmoMC \cite{Lewis:2002ah}. 
We consider an extended dataset comprising Cosmic Microwave Background measurements, through the Planck (2018) likelihoods \cite{Planck:2019nip}  \footnote{ We use Plik likelihood  ``TT,TE,EE+lowE" by combination of temperature TT, polarization EE and their cross-correlation TE power spectra  over the range $\ell \in [30, 2508]$, the low-$\ell$ temperature Commander likelihood, and the low-$\ell$ SimAll EE likelihood.}, the CMB lensing reconstruction power spectrum~\cite{Planck:2019nip,Planck:2018lbu}, the Baryon Acoustic Oscillation measurements from 6dFGS~\cite{Beutler:2011hx}, SDSS-MGS~\cite{Ross:2014qpa}, and BOSS DR12~\cite{BOSS:2016wmc} surveys, the type Ia SNe Pantheon compilation~\cite{Scolnic:2017caz} and cosmic chronometers measurements of the expansion rate $H(z)$ from the relative ages of passively evolving galaxies \cite{Jimenez:2001gg,Stern:2009ep,Moresco:2015cya,Zhang:2012mp,Moresco:2016mzx,Ratsimbazafy:2017vga}.  We also consider a Gaussian prior for the SNe Ia absolute magnitude $M$, in order to consider the calibration given by the current $H_0$ local measurements \cite{Riess:2016jrr}, $(-19.2435 \pm 0.0373)$ mag, as suggested in \cite{Camarena:2021jlr}. This approach prevents double counting of low redshift supernovae and avoids to assume a value of the deceleration parameter, considering $M$ constrained by the local calibration of SNeI, which is not included otherwise.
Let we stress that several approaches are been considered in the literature to   address the   so called  `$H_0$ tension'.
The initial approaches started by  fixing  the value of $H_0$, or imposing a prior on it in the statistical analysis \cite{Benetti:2017gvm,Benetti:2017juy}, but the method proved to be statistically inadequate \cite{Efstathiou:2020wem,Gonzalez:2021ojp,vagnozzi:2020dfn,Moresco:2022phi} as it attempts to combine two data sets that are intrinsically incompatible when considered in the context of the $\Lambda$CDM model,  i.e. CMB and SNeIa, and limits the parametric space of the analysis in regions not determined   by the CMB data alone. Thus, it has been shown that it is statistically more acceptable to consider a prior on the supernovae calibration parameter instead \cite{Camarena:2021jlr}. Of course, it is crucial to test if this incompatibility of supernovae and CMB data sets still holds in the context of the unimodular model and this will be analyzed in the Results section.

In our analysis, we vary the usual cosmological parameters, namely, the physical baryon density, $\Omega_B h^2$, the physical
cold dark matter density, $\Omega_{DM} h^2$, the ratio between the sound horizon and the angular diameter distance at decoupling,
$\theta$, the optical depth $\tau$, the primordial amplitude, $A_s$ and the spectral index $n_s$. We also vary the unimodular model parameters $\Delta \rho_\Lambda$, $a^*$ and $\delta$ and the nuisance foreground parameters \cite{2016A&A...594A..11P}. We assume large flat priors for the free parameters, vaying the unimodular ones between $\Delta \rho_\Lambda \in [-0.2;0.2]$ , $a^* \in [0.02;1]$ and $\delta \in [0.02;1]$ .
We consider purely adiabatic initial conditions. The
sum of neutrino masses is fixed to 0.06 eV, and we limit the analysis to scalar perturbations with $k^*=0.05$ Mpc. 

We choose to perform two statistical analyses that we describe as follows: i) the \textit{main} analysis, which we simply call ``Unimodular parameters free", in which we let both the cosmological and the parameters of the unimodular model free to vary , ii) a \textit{speculative} analysis, which we refer to as ``Unimodular parameters fixed"   where we fix the value of the parameters of the unimodular model to arbitrary values and let the other cosmological parameters vary. 
This second choice is due to the fact that, as mentioned in the Section \ref{theory2},  a degeneration between $\Delta \rho_{\Lambda}$ and the cosmological parameters  $\Omega_{DM}$, $\Omega_B$ and $H_0$ is expected and will be shown next in the results section. 
Such degeneration can allow this model to predict higher $H_0$ values, and here we want to show the cost of this achievement and the ability of this model to alleviate the Hubble tension whether future data may show better sensitivity to the constraint of its parameters.


\begin{table}[]
\centering
\caption{{
Analysis constraints for the models parameters using the data set Planck(2018)+lensing+BAO+Pantheon+CC+M prior. Quoted intervals correspond to 68\%~C.L. intervals, whereas quoted upper/lower limits correspond to 95\%~C.L. upper/lower limits.
}
\label{tab:results}}
\begin{tabular}{|c|c|c|c|c|c|c|}
\hline
\multicolumn{1}{|c|}{Parameter}&
\multicolumn{2}{c}{$\Lambda$CDM}&
\multicolumn{2}{|c|}{Unimodular  parameters free}&
\multicolumn{2}{|c|}{Unimodular  parameters fixed}\\
\hline
{ }&
{mean value and }&
{bestfit}&
{mean value and }&
{bestfit}&
{mean value and}&
{bestfit}\\
 & $68 \%$ confidence levels & & $68 \%$ confidence levels & &  $68 \%$ confidence levels & \\
\hline
$100\,\Omega_B h^2$ 	
& $2.250 \pm 0.013$ %
& $2.254$ %
& $2.242 \pm 0.027$ %
& $2.257$ %
& $2.123 \pm 0.012$ %
& $2.132$ %
\\
$\Omega_{\rm DM} h^2$	
& $0.1186 \pm 0.0009$ %
& $0.1191$ %
& $0.1181 \pm 0.0017$ %
& $0.1189$ 
& $0.1140 \pm 0.0009$ %
& $0.1134$ %
\\
$\tau$	
& $0.058 \pm 0.007$ %
& $0.053$ %
& $0.056 \pm 0.007$ 
& $0.060$ 
& $0.053 \pm 0.007$ %
& $0.052$ %
\\
${\rm{ln}}(10^{10} A_s)$ 	
& $3.050 \pm 0.015$ %
& $3.040$ %
& $3.045 \pm 0.015$ %
& $3.055$ %
& $3.058 \pm 0.014$ %
& $3.056$ %
\\
$n_s$	
& $0.9684 \pm 0.0036$ %
& $0.9688$ %
& $0.9700 \pm 0.0056$ %
& $0.966$ 
& $0.984 \pm 0.004$ %
& $0.982$ %
\\
$\Delta {\rho}_{\Lambda}$	
& $-$ %
& $-$ %
& $0.0010 \pm 0.0040$ %
& $-0.0003$ 
& fixed to $0.09 $ 
& $-$ %
\\
$a^*$	
& $-$ %
& $-$ %
& $>0.4$ %
& $0.8$ 
& fixed to $0.4$  
& $-$ %
\\
$\delta$	
& $-$ %
& $-$ %
& $<0.65$ %
& $0.22$ 
& fixed to $0.02$  
& $-$ %
\\
$H_0$ [Km/s/Mpc]
& $68.02 \pm 0.40$ %
& $67.88$ %
& $ 68.51 \pm 0.57 $ 
& $ 68.24$ %
& $71.18 \pm 0.42$ %
& $71.43$ %
\\
$100 \Omega_B^{\rm early} h^2$
& $-$ %
& $-$ %
& $ 2.245^{+ 0.038}_{-0.035} $ 
& $ 2.255$ %
& $2.173 \pm 0.024$ %
& $2.182$ %
\\
$ \Omega_{DM}^{\rm early} h^2$
& $-$ %
& $-$ %
& $ 0.1190 \pm 0.0025 $ 
& $ 0.1195 $ %
& $0.1174 \pm 0.0018$ %
& $0.1168$ %
\\
\hline
\end{tabular}
\end{table} 
\section{Results}
\label{results}
We present the results our statistical analyses in Tab. \ref{tab:results} and Figs. \ref{fig:tri_cosmo}, \ref{fig:tri_omegas} and \ref{fig:tri_unimodular}. Note that we also included the results of a $\Lambda$CDM model analysis using the same data set for comparison. About our main analysis, i.e. when the unimodular parameters are free to vary, we found an agreement within 1$\sigma$  with the $\Lambda$CDM model cosmological parameters, as shown in the second column of the table and with grey curves in the plots.
At the same time, both $\Delta \rho_\Lambda$ and $\delta$ are well  constrained (see left panel of Fig.\ref{fig:tri_unimodular}) while just a lower bound on $a^*$ can be  established. Here we stress that the data clearly indicate that the anomalous behaviour in the energy densities must take place much later than the formation of neutral hydrogen ($a_{rec}\sim 10^{-3}$).  

The $H_0$ value of this analyses is compatible with that of the $\Lambda$CDM, although it should be noted that slightly higher values of $H_0$ are allowed. 
As expected (see discussion in Section \ref{Cl}),  $\Delta \rho_{\Lambda}$ shows degeneracy with the matter densities values and a weak correlation with $H_0$, as shown in the right panel of Fig. \ref{fig:tri_unimodular}. On the other hand, the $a^*$ and $\delta$ parameters show no degeneracy with the cosmological parameters, which can be explained by the low sensitivity of the CMB spectrum on these parameters (see discussion in section \ref{Cl}).  
On the other hand, we recall that in our  model, the cosmological constant takes two different fixed values over different epochs   of  standard
 cosmological evolution (see equations \ref{DE_2} and \ref{fa}).
  Our results for  $\Delta \rho_\Lambda$ show that the difference between them is small, namely of order $0.33\%$. 
  Finally, it is worth mentioning that the $\chi^2_{\min}$ of this model is comparable to that of the standard model,
   but has three more parameters in the theory. 
   On the other hand, one of the parameters of the unimodular model is not well constrained and this deserves further investigation with
      next generation of data that may be more sensitive to the
      modifications to the usual  theory analyzed here.

Given the difficulty in constraining the $a^*$  parameter, and the apparent lack of degeneracy of $\delta$ with cosmological parameters, we now focus on  a model where both $a^*$ and $\delta$ parameter are fixed  at the smallest values allowed by our first analysis, $0.4$ and  $0.02$ respectively. Besides,  the value of $\Delta \rho_{\Lambda}$ is set at a large positive arbitrary value which lies within the $1\sigma$ confidence interval obtained in  the first analysis (i.e where all parameters varying freely). This allows us to perform a second speculative analysis, shown in red in the Figs. \ref{fig:tri_cosmo}, \ref{fig:tri_omegas} and in the last columns of the Tab. \ref{tab:results}. We found that for this choice of unimodular parameter values, the $H_0$ is shifted into higher values with respect to the $\Lambda$CDM model, in agreement within 1.9$\sigma$  with the latest value of Riess et al.  \cite{2019ApJ...876...85R} obtained from local measurements. 
 This is an important result of the present work, because it provides a guide for the construction of models that describe the density of black holes, which are, according to our model, the scenario where the violation of the conservation of the tensor-energy moment occurs and therefore, gives rise to the distinctive behavior of the model. Our results show that higher values of $\Delta \rho$ shift  $H_0$ to values that further alleviate the Hubble tension and therefore, the models that describe the energy diffusion in black holes (which are at present under construction)  should, in order to fully solve the issue,  point toward these values. We recall that in the present work, we assumed a simple model for the behaviour of $\rho_\Lambda$, which is supposed to arise from an energy diffussion from matter (both dark and baryonic) to dark energy catalyzed by black holes. 
Likewise the predicted values of $n_s$ are also not in agreement with those of the $\Lambda$CDM model, but still consistent with the predictions of standard inflationary models. About the constraints for $\Omega_B h^2$ and $\Omega_{DM} h^2$, there is also not agreement within 2$\sigma$ with the $\Lambda$CDM model. However, as discussed in section \ref{theory2}, the behavior of these quantities is not the same in both models and therefore, we do not expect to obtain the same estimation. Indeed, for the model to be viable it is necessary that the predicted values of the baryon density in the early universe $\Omega_b^{\rm early} h^2$ are consistent with  the nucleosynthesis constraint $\Omega_b h^2=(0.021,0.024)$ and this is verified by the results shown in Tab. \ref{tab:results}.

Finally, we use the Deviance Information Criterion (DIC)~\cite{DIC} in order   to test whether the complexity of UG model is statistically supported by the data with respect to the vanilla $\Lambda$CDM one. The DIC  has  been proven to be a useful tool to test the average performance of a model with a penalty given by the Bayesian complexity ~\cite{Kunz:2006mc,Trotta:2008qt,Spiegelhalter:2002yvw} (see \cite{SantosdaCosta:2020dyl,Frusciante:2020gkx,Winkler:2019hkh} and references inside for some applications of DIC in cosmology). The DIC value is defined, for the selected model $\mathcal{M}$, as:
\begin{equation}\label{eq:dic}
DIC_\mathcal{M}\equiv-2 {\ln\mathcal{L}(\theta)}+2p_D\,,
\end{equation}
where the first term is the posterior mean of $\mathcal{L}(\theta)$, i.e. the likelihood of the data given the model parameters $\theta$, and the second term is the Bayesian complexity $p_D=-2\overline{\ln\mathcal{L}(\theta)}+2\ln\mathcal{L}(\tilde{\theta})$ where the tilde refers  to the chosen estimator.
\footnote{ We choose to use the best fit as estimator, so that the DIC can be rewritten as $DIC_\mathcal{M}=2 {\ln\mathcal{L}({\theta})}-4\overline{\ln\mathcal{L}(\theta)}$. We obtain the mean likelihood from the output chains of the MCMC analysis, and the best fit likelihood via the BOBYQA algorithm implemented in CosmoMC for likelihood maximisation.}  
We considered following  \cite{Kass:1995loi} the following scale to  determine the performance of  our model  in that  test: $\Delta DIC=10/5/1$  as indicating, respectively, strong/moderate/null preference for the reference model (if the value of $\Delta DIC$ is negative,  it would indicate a preference for the model under consideration). 
As we obtained $\Delta$DIC $=3.6$., we conclude that there is no evidence, according to this  criteria, to  support  the analysed UG model over  the standard $\Lambda$CDM one. This indicates  that the current data are not sensitive enough to detect the full complexity of the model. However, we stress that  even using  this   strict  measure the  model is not discarded  in comparison to  $\Lambda$CDM.

\section{Summary and Conclusions}
\label{conclusions}

  In this work, we perform a methodologically  proper analysis on a general  idea involving     late time violation of energy-momentum  conservation,   resulting from a  kind of  granularity    of space-time   whose ultimate origin lies in quantum gravitational features. For this, it is necessary to formulate the cosmological model in the context of Unimodular Gravity,  a  modification of  GR  that allows  such  violations   (under  certain conditions that  hold   automatically  in cosmology).   This idea was applied to the very early universe to offer  an  account of the  nature  and   magnitude  of the cosmological constant,  a fact that  serves  as  a strong motivation  to  seek a resolution of the $H_0$ tension  on  similar grounds \cite{Josset:2016vrq,SP2019}. 
  
  The analysis carried  out in  this work must however be considered as  preliminary since the basic idea is that the effect would be linked to a kind of
  effective  friction   affecting  black holes  in the   epoch  between   last  scattering and the present.  However, we  lack, at this  time,  a clear  picture of the   black  hole   abundances  as a function of   their mass   angular momentum and time    in  the relevant period. This   forces us to   use  a  rather  simple   and  rudimentary  model of   the   effect in terms  of an  anomalous   evolution of  the   energy budget  during the relevant epochs.
 
 We analyse the model using a Boltzmann solver code to take into account both background and perturbation evolution. By comparison with a selected set of data, such as CMB, BAO, SnIa and CC, we are able to constrain, to the best of the current ability, the free parameters of the theory. In addition, we note the sensitivity limits of the data on constraining model complexity and discuss in depth the degeneracies between parameters. 
 In summary, we are only able to constrain one unimodular parameter, $\Delta \rho_{\Lambda}$, while we can at most find upper/lower bounds for the other two, $a^*$ and $\delta$. We do not notice significant changes on the constraint of cosmological parameters compared to the standard cosmological model.

 It is  however  a  noteworthy  fact , that  our results show that the model does not spontaneously reduce the tension on $H_0$, producing only a small shift in the value of the parameter. However, it must be emphasised that the potential of the model is not fully expressed as the data proved unable to full constrain the unimodular parameters. In fact, looking at the results of our speculative analysis (in which we fix the values of the three unimodular parameters at arbitrarily chosen values) we see the $H_0$ tension being  relaxed, resulting in  the value of $H_0=71.6$ Km/s/Mpc at $1 \sigma$. This implies
 very   small modification of most other   cosmological parameters  and  without leading to  an  a problematic  depletion of the  dark  matter and  baryonic  components  (dark and   bright)   which are  of course required  at late times to  account for  the present   features  of   galaxies  and   galactic  clusters.  More  specifically the  present day  energy  budget   resulting from the  model is the following  ranges (the following values are at $68\%$ confidence level):  $\Omega_B=(0.0412,0.0543)$,  $\Omega_{DM}=(0.2438,0.2594)$, $\Omega_{\Lambda}=(0.69,0.71)$ when all parameters are free to vary and $\Omega_B=(0.0453,0.0458)$,  $\Omega_{DM}=(0.2429,0.2467)$,  $\Omega_{\Lambda}=(0.73,0.74)$ when the unimodular parameters are fixed  which    can be compared  with the  standard     $ \Lambda {\rm CDM}$   values given by  $\Omega_B=(0.0505,0.0574)$, $\Omega_{DM}=(0.2513,0.2613)$, $\Omega_{\Lambda}=(0.69,0.70)$

  In our  view this   clearly illustrates the  potential of the proposal  to   fully resolve the   $ H_0$  tension,   once  a    more clear  picture  of the detail form of the function $ f(M, J,t)$ in  known  (or  realistically  characterized in terms of a  suitable  set of parameters) allowing a repetition  on the analysis   carried out 
  in  this work  with  better modeling  of the relevant  black hole abundances.

Finally, since the predictions of the unimodular model for the B modes are different from the ones of the standard model, future CMB polarization data should provide more strict constraints on the unimodular model.

\section{Aknowledgements}
 
 SL is supported by grant PIP 11220200100729CO CONICET and grant 20020170100129BA UBACYT.
 MB acknowledge Istituto Nazionale di Fisica Nucleare (INFN), sezione di Napoli, \textit{iniziativa specifica} QGSKY. 
 DS acknowledges partial financial support from  PAPIIT-DGAPA-UNAM project IG100120 and   CONACyT project 140630.   He is grateful for the support provided by the grant FQXI-MGA-1920 from the Foundational Questions Institute and the Fetzer Franklin Fund, a donor advised by the Silicon Valley Community Foundation.  
 
 The authors acknowledge the use of the supercluster-
MIZTLI of UNAM through project LANCAD-UNAM-
DGTIC-132 and thank the people of DGTIC-UNAM for
technical and computational support. Also, we thank the use of CosmoMC code, and also acknowledge the use of the High Performance Data Center (DCON) at the Observatorio Nacional for providing the computational facilities to run our analysis. 
 
\bibliographystyle{apsrev}
\bibliography{bibliografia3.bib}

\begin{thebibliography}{60}
\expandafter\ifx\csname natexlab\endcsname\relax\def\natexlab#1{#1}\fi
\expandafter\ifx\csname bibnamefont\endcsname\relax
  \def\bibnamefont#1{#1}\fi
\expandafter\ifx\csname bibfnamefont\endcsname\relax
  \def\bibfnamefont#1{#1}\fi
\expandafter\ifx\csname citenamefont\endcsname\relax
  \def\citenamefont#1{#1}\fi
\expandafter\ifx\csname url\endcsname\relax
  \def\url#1{\texttt{#1}}\fi
\expandafter\ifx\csname urlprefix\endcsname\relax\def\urlprefix{URL }\fi
\providecommand{\bibinfo}[2]{#2}
\providecommand{\eprint}[2][]{\url{#2}}

\bibitem[{\citenamefont{Maudlin et~al.}(2020)\citenamefont{Maudlin, Okon, and
  Sudarsky}}]{Maudlin:2019bje}
\bibinfo{author}{\bibfnamefont{T.}~\bibnamefont{Maudlin}},
  \bibinfo{author}{\bibfnamefont{E.}~\bibnamefont{Okon}}, \bibnamefont{and}
  \bibinfo{author}{\bibfnamefont{D.}~\bibnamefont{Sudarsky}},
  \bibinfo{journal}{Stud. Hist. Phil. Sci. B} \textbf{\bibinfo{volume}{69}},
  \bibinfo{pages}{67} (\bibinfo{year}{2020}), \eprint{1910.06473}.

\bibitem[{\citenamefont{Josset et~al.}(2017)\citenamefont{Josset, Perez, and
  Sudarsky}}]{Josset:2016vrq}
\bibinfo{author}{\bibfnamefont{T.}~\bibnamefont{Josset}},
  \bibinfo{author}{\bibfnamefont{A.}~\bibnamefont{Perez}}, \bibnamefont{and}
  \bibinfo{author}{\bibfnamefont{D.}~\bibnamefont{Sudarsky}},
  \bibinfo{journal}{Phys. Rev. Lett.} \textbf{\bibinfo{volume}{118}},
  \bibinfo{pages}{021102} (\bibinfo{year}{2017}), \eprint{1604.04183}.

\bibitem[{\citenamefont{{Wald}}(1984)}]{Wald:1984}
\bibinfo{author}{\bibfnamefont{R.~M.} \bibnamefont{{Wald}}},
  \emph{\bibinfo{title}{{General Relativity}}} (\bibinfo{year}{1984}).

\bibitem[{\citenamefont{Perez et~al.}(2018)\citenamefont{Perez, Sudarsky, and
  Bjorken}}]{Perez:2018wlo}
\bibinfo{author}{\bibfnamefont{A.}~\bibnamefont{Perez}},
  \bibinfo{author}{\bibfnamefont{D.}~\bibnamefont{Sudarsky}}, \bibnamefont{and}
  \bibinfo{author}{\bibfnamefont{J.~D.} \bibnamefont{Bjorken}},
  \bibinfo{journal}{Int. J. Mod. Phys. D} \textbf{\bibinfo{volume}{27}},
  \bibinfo{pages}{1846002} (\bibinfo{year}{2018}), \eprint{1804.07162}.

\bibitem[{\citenamefont{{Perez} and {Sudarsky}}(2019)}]{SP2019}
\bibinfo{author}{\bibfnamefont{A.}~\bibnamefont{{Perez}}} \bibnamefont{and}
  \bibinfo{author}{\bibfnamefont{D.}~\bibnamefont{{Sudarsky}}},
  \bibinfo{journal}{\prl} \textbf{\bibinfo{volume}{122}}, \bibinfo{eid}{221302}
  (\bibinfo{year}{2019}).

\bibitem[{\citenamefont{Amadei and Perez}(2022)}]{Amadei:2021aqd}
\bibinfo{author}{\bibfnamefont{L.}~\bibnamefont{Amadei}} \bibnamefont{and}
  \bibinfo{author}{\bibfnamefont{A.}~\bibnamefont{Perez}},
  \bibinfo{journal}{Phys. Rev. D} \textbf{\bibinfo{volume}{106}},
  \bibinfo{pages}{063528} (\bibinfo{year}{2022}), \eprint{2104.08881}.

\bibitem[{\citenamefont{Perez and Sudarsky}(2019)}]{Perez:2019gyd}
\bibinfo{author}{\bibfnamefont{A.}~\bibnamefont{Perez}} \bibnamefont{and}
  \bibinfo{author}{\bibfnamefont{D.}~\bibnamefont{Sudarsky}}
  (\bibinfo{year}{2019}), \eprint{1911.06059}.

\bibitem[{\citenamefont{Perez et~al.}(2021)\citenamefont{Perez, Sudarsky, and
  Wilson-Ewing}}]{Perez:2020cwa}
\bibinfo{author}{\bibfnamefont{A.}~\bibnamefont{Perez}},
  \bibinfo{author}{\bibfnamefont{D.}~\bibnamefont{Sudarsky}}, \bibnamefont{and}
  \bibinfo{author}{\bibfnamefont{E.}~\bibnamefont{Wilson-Ewing}},
  \bibinfo{journal}{Gen. Rel. Grav.} \textbf{\bibinfo{volume}{53}},
  \bibinfo{pages}{7} (\bibinfo{year}{2021}), \eprint{2001.07536}.

\bibitem[{\citenamefont{{Planck Collaboration}
  et~al.}(2020)\citenamefont{{Planck Collaboration}, {Aghanim}, {Akrami},
  {Ashdown}, {Aumont}, {Baccigalupi}, {Ballardini}, {Banday}, {Barreiro},
  {Bartolo} et~al.}}]{planckcosmoparams18}
\bibinfo{author}{\bibnamefont{{Planck Collaboration}}},
  \bibinfo{author}{\bibfnamefont{N.}~\bibnamefont{{Aghanim}}},
  \bibinfo{author}{\bibfnamefont{Y.}~\bibnamefont{{Akrami}}},
  \bibinfo{author}{\bibfnamefont{M.}~\bibnamefont{{Ashdown}}},
  \bibinfo{author}{\bibfnamefont{J.}~\bibnamefont{{Aumont}}},
  \bibinfo{author}{\bibfnamefont{C.}~\bibnamefont{{Baccigalupi}}},
  \bibinfo{author}{\bibfnamefont{M.}~\bibnamefont{{Ballardini}}},
  \bibinfo{author}{\bibfnamefont{A.~J.} \bibnamefont{{Banday}}},
  \bibinfo{author}{\bibfnamefont{R.~B.} \bibnamefont{{Barreiro}}},
  \bibinfo{author}{\bibfnamefont{N.}~\bibnamefont{{Bartolo}}},
  \bibnamefont{et~al.}, \bibinfo{journal}{Astronomy and Astrophysics}
  \textbf{\bibinfo{volume}{641}}, \bibinfo{eid}{A6} (\bibinfo{year}{2020}),
  \eprint{1807.06209}.

\bibitem[{\citenamefont{{Riess} et~al.}(2019)\citenamefont{{Riess},
  {Casertano}, {Yuan}, {Macri}, and {Scolnic}}}]{2019ApJ...876...85R}
\bibinfo{author}{\bibfnamefont{A.~G.} \bibnamefont{{Riess}}},
  \bibinfo{author}{\bibfnamefont{S.}~\bibnamefont{{Casertano}}},
  \bibinfo{author}{\bibfnamefont{W.}~\bibnamefont{{Yuan}}},
  \bibinfo{author}{\bibfnamefont{L.~M.} \bibnamefont{{Macri}}},
  \bibnamefont{and}
  \bibinfo{author}{\bibfnamefont{D.}~\bibnamefont{{Scolnic}}},
  \bibinfo{journal}{Astrophysical Journal} \textbf{\bibinfo{volume}{876}},
  \bibinfo{eid}{85} (\bibinfo{year}{2019}), \eprint{1903.07603}.

\bibitem[{\citenamefont{{Riess} et~al.}(2021)\citenamefont{{Riess}, {Yuan},
  {Macri}, {Scolnic}, {Brout}, {Casertano}, {Jones}, {Murakami}, {Breuval},
  {Brink} et~al.}}]{2021arXiv211204510R}
\bibinfo{author}{\bibfnamefont{A.~G.} \bibnamefont{{Riess}}},
  \bibinfo{author}{\bibfnamefont{W.}~\bibnamefont{{Yuan}}},
  \bibinfo{author}{\bibfnamefont{L.~M.} \bibnamefont{{Macri}}},
  \bibinfo{author}{\bibfnamefont{D.}~\bibnamefont{{Scolnic}}},
  \bibinfo{author}{\bibfnamefont{D.}~\bibnamefont{{Brout}}},
  \bibinfo{author}{\bibfnamefont{S.}~\bibnamefont{{Casertano}}},
  \bibinfo{author}{\bibfnamefont{D.~O.} \bibnamefont{{Jones}}},
  \bibinfo{author}{\bibfnamefont{Y.}~\bibnamefont{{Murakami}}},
  \bibinfo{author}{\bibfnamefont{L.}~\bibnamefont{{Breuval}}},
  \bibinfo{author}{\bibfnamefont{T.~G.} \bibnamefont{{Brink}}},
  \bibnamefont{et~al.}, \bibinfo{journal}{arXiv e-prints}
  \bibinfo{eid}{arXiv:2112.04510} (\bibinfo{year}{2021}), \eprint{2112.04510}.

\bibitem[{\citenamefont{{Linares Cede{\~n}o} and
  {Nucamendi}}(2021)}]{2021PDU....3200807L}
\bibinfo{author}{\bibfnamefont{F.~X.} \bibnamefont{{Linares Cede{\~n}o}}}
  \bibnamefont{and}
  \bibinfo{author}{\bibfnamefont{U.}~\bibnamefont{{Nucamendi}}},
  \bibinfo{journal}{Physics of the Dark Universe}
  \textbf{\bibinfo{volume}{32}}, \bibinfo{eid}{100807} (\bibinfo{year}{2021}),
  \eprint{2009.10268}.

\bibitem[{\citenamefont{{Akarsu} et~al.}(2021)\citenamefont{{Akarsu}, {Kumar},
  {{\"O}z{\"u}lker}, and {Vazquez}}}]{2021PhRvD.104l3512A}
\bibinfo{author}{\bibfnamefont{{\"O}.}~\bibnamefont{{Akarsu}}},
  \bibinfo{author}{\bibfnamefont{S.}~\bibnamefont{{Kumar}}},
  \bibinfo{author}{\bibfnamefont{E.}~\bibnamefont{{{\"O}z{\"u}lker}}},
  \bibnamefont{and} \bibinfo{author}{\bibfnamefont{J.~A.}
  \bibnamefont{{Vazquez}}}, \bibinfo{journal}{\prd}
  \textbf{\bibinfo{volume}{104}}, \bibinfo{eid}{123512} (\bibinfo{year}{2021}),
  \eprint{2108.09239}.

\bibitem[{\citenamefont{{Carr} et~al.}(2021)\citenamefont{{Carr}, {Kohri},
  {Sendouda}, and {Yokoyama}}}]{2021RPPh...84k6902C}
\bibinfo{author}{\bibfnamefont{B.}~\bibnamefont{{Carr}}},
  \bibinfo{author}{\bibfnamefont{K.}~\bibnamefont{{Kohri}}},
  \bibinfo{author}{\bibfnamefont{Y.}~\bibnamefont{{Sendouda}}},
  \bibnamefont{and}
  \bibinfo{author}{\bibfnamefont{J.}~\bibnamefont{{Yokoyama}}},
  \bibinfo{journal}{Reports on Progress in Physics}
  \textbf{\bibinfo{volume}{84}}, \bibinfo{eid}{116902} (\bibinfo{year}{2021}),
  \eprint{2002.12778}.

\bibitem[{\citenamefont{{Morikawa}}(2017)}]{2017EPJWC.16401011M}
\bibinfo{author}{\bibfnamefont{M.}~\bibnamefont{{Morikawa}}}, in
  \emph{\bibinfo{booktitle}{European Physical Journal Web of Conferences}}
  (\bibinfo{year}{2017}), vol. \bibinfo{volume}{164} of
  \emph{\bibinfo{series}{European Physical Journal Web of Conferences}}, p.
  \bibinfo{pages}{01011}.

\bibitem[{\citenamefont{{Green} and {Kavanagh}}(2021)}]{2021JPhG...48d3001G}
\bibinfo{author}{\bibfnamefont{A.~M.} \bibnamefont{{Green}}} \bibnamefont{and}
  \bibinfo{author}{\bibfnamefont{B.~J.} \bibnamefont{{Kavanagh}}},
  \bibinfo{journal}{Journal of Physics G Nuclear Physics}
  \textbf{\bibinfo{volume}{48}}, \bibinfo{eid}{043001} (\bibinfo{year}{2021}),
  \eprint{2007.10722}.

\bibitem[{\citenamefont{{Carr} et~al.}(2010)\citenamefont{{Carr}, {Kohri},
  {Sendouda}, and {Yokoyama}}}]{2010PhRvD..81j4019C}
\bibinfo{author}{\bibfnamefont{B.~J.} \bibnamefont{{Carr}}},
  \bibinfo{author}{\bibfnamefont{K.}~\bibnamefont{{Kohri}}},
  \bibinfo{author}{\bibfnamefont{Y.}~\bibnamefont{{Sendouda}}},
  \bibnamefont{and}
  \bibinfo{author}{\bibfnamefont{J.}~\bibnamefont{{Yokoyama}}},
  \bibinfo{journal}{\prd} \textbf{\bibinfo{volume}{81}}, \bibinfo{eid}{104019}
  (\bibinfo{year}{2010}), \eprint{0912.5297}.

\bibitem[{\citenamefont{{Niikura} et~al.}(2019)\citenamefont{{Niikura},
  {Takada}, {Yasuda}, {Lupton}, {Sumi}, {More}, {Kurita}, {Sugiyama}, {More},
  {Oguri} et~al.}}]{2019NatAs...3..524N}
\bibinfo{author}{\bibfnamefont{H.}~\bibnamefont{{Niikura}}},
  \bibinfo{author}{\bibfnamefont{M.}~\bibnamefont{{Takada}}},
  \bibinfo{author}{\bibfnamefont{N.}~\bibnamefont{{Yasuda}}},
  \bibinfo{author}{\bibfnamefont{R.~H.} \bibnamefont{{Lupton}}},
  \bibinfo{author}{\bibfnamefont{T.}~\bibnamefont{{Sumi}}},
  \bibinfo{author}{\bibfnamefont{S.}~\bibnamefont{{More}}},
  \bibinfo{author}{\bibfnamefont{T.}~\bibnamefont{{Kurita}}},
  \bibinfo{author}{\bibfnamefont{S.}~\bibnamefont{{Sugiyama}}},
  \bibinfo{author}{\bibfnamefont{A.}~\bibnamefont{{More}}},
  \bibinfo{author}{\bibfnamefont{M.}~\bibnamefont{{Oguri}}},
  \bibnamefont{et~al.}, \bibinfo{journal}{Nature Astronomy}
  \textbf{\bibinfo{volume}{3}}, \bibinfo{pages}{524} (\bibinfo{year}{2019}),
  \eprint{1701.02151}.

\bibitem[{\citenamefont{{Hawkins}}(1993)}]{1993Natur.366..242H}
\bibinfo{author}{\bibfnamefont{M.~R.~S.} \bibnamefont{{Hawkins}}},
  \bibinfo{journal}{\nat} \textbf{\bibinfo{volume}{366}}, \bibinfo{pages}{242}
  (\bibinfo{year}{1993}).

\bibitem[{\citenamefont{{Zumalac{\'a}rregui} and
  {Seljak}}(2018)}]{2018PhRvL.121n1101Z}
\bibinfo{author}{\bibfnamefont{M.}~\bibnamefont{{Zumalac{\'a}rregui}}}
  \bibnamefont{and} \bibinfo{author}{\bibfnamefont{U.}~\bibnamefont{{Seljak}}},
  \bibinfo{journal}{\prl} \textbf{\bibinfo{volume}{121}}, \bibinfo{eid}{141101}
  (\bibinfo{year}{2018}), \eprint{1712.02240}.

\bibitem[{\citenamefont{{Mack} et~al.}(2007)\citenamefont{{Mack}, {Ostriker},
  and {Ricotti}}}]{2007ApJ...665.1277M}
\bibinfo{author}{\bibfnamefont{K.~J.} \bibnamefont{{Mack}}},
  \bibinfo{author}{\bibfnamefont{J.~P.} \bibnamefont{{Ostriker}}},
  \bibnamefont{and}
  \bibinfo{author}{\bibfnamefont{M.}~\bibnamefont{{Ricotti}}},
  \bibinfo{journal}{\apj} \textbf{\bibinfo{volume}{665}}, \bibinfo{pages}{1277}
  (\bibinfo{year}{2007}), \eprint{astro-ph/0608642}.

\bibitem[{\citenamefont{{Kohri} et~al.}(2014)\citenamefont{{Kohri}, {Nakama},
  and {Suyama}}}]{2014PhRvD..90h3514K}
\bibinfo{author}{\bibfnamefont{K.}~\bibnamefont{{Kohri}}},
  \bibinfo{author}{\bibfnamefont{T.}~\bibnamefont{{Nakama}}}, \bibnamefont{and}
  \bibinfo{author}{\bibfnamefont{T.}~\bibnamefont{{Suyama}}},
  \bibinfo{journal}{\prd} \textbf{\bibinfo{volume}{90}}, \bibinfo{eid}{083514}
  (\bibinfo{year}{2014}), \eprint{1405.5999}.

\bibitem[{\citenamefont{{Capela} et~al.}(2013)\citenamefont{{Capela},
  {Pshirkov}, and {Tinyakov}}}]{2013PhRvD..87l3524C}
\bibinfo{author}{\bibfnamefont{F.}~\bibnamefont{{Capela}}},
  \bibinfo{author}{\bibfnamefont{M.}~\bibnamefont{{Pshirkov}}},
  \bibnamefont{and}
  \bibinfo{author}{\bibfnamefont{P.}~\bibnamefont{{Tinyakov}}},
  \bibinfo{journal}{\prd} \textbf{\bibinfo{volume}{87}}, \bibinfo{eid}{123524}
  (\bibinfo{year}{2013}), \eprint{1301.4984}.

\bibitem[{\citenamefont{Collins et~al.}(2004)\citenamefont{Collins, Perez,
  Sudarsky, Urrutia, and Vucetich}}]{Collins-2004}
\bibinfo{author}{\bibfnamefont{J.}~\bibnamefont{Collins}},
  \bibinfo{author}{\bibfnamefont{A.}~\bibnamefont{Perez}},
  \bibinfo{author}{\bibfnamefont{D.}~\bibnamefont{Sudarsky}},
  \bibinfo{author}{\bibfnamefont{L.}~\bibnamefont{Urrutia}}, \bibnamefont{and}
  \bibinfo{author}{\bibfnamefont{H.}~\bibnamefont{Vucetich}},
  \bibinfo{journal}{Phys. Rev. Lett.} \textbf{\bibinfo{volume}{93}},
  \bibinfo{pages}{191301} (\bibinfo{year}{2004}),
  \urlprefix\url{https://link.aps.org/doi/10.1103/PhysRevLett.93.191301}.

\bibitem[{\citenamefont{{Perez} et~al.}(2021)\citenamefont{{Perez}, {Sudarsky},
  and {Wilson-Ewing}}}]{PS2020}
\bibinfo{author}{\bibfnamefont{A.}~\bibnamefont{{Perez}}},
  \bibinfo{author}{\bibfnamefont{D.}~\bibnamefont{{Sudarsky}}},
  \bibnamefont{and}
  \bibinfo{author}{\bibfnamefont{E.}~\bibnamefont{{Wilson-Ewing}}},
  \bibinfo{journal}{General Relativity and Gravitation}
  \textbf{\bibinfo{volume}{53}}, \bibinfo{eid}{7} (\bibinfo{year}{2021}),
  \eprint{2001.07536}.

\bibitem[{\citenamefont{{Unruh}}(1989)}]{1989PhRvD..40.1048U}
\bibinfo{author}{\bibfnamefont{W.~G.} \bibnamefont{{Unruh}}},
  \bibinfo{journal}{\prd} \textbf{\bibinfo{volume}{40}}, \bibinfo{pages}{1048}
  (\bibinfo{year}{1989}).

\bibitem[{\citenamefont{{Weinberg}}(1989)}]{1989RvMP...61....1W}
\bibinfo{author}{\bibfnamefont{S.}~\bibnamefont{{Weinberg}}},
  \bibinfo{journal}{Reviews of Modern Physics} \textbf{\bibinfo{volume}{61}},
  \bibinfo{pages}{1} (\bibinfo{year}{1989}).

\bibitem[{\citenamefont{Kostelecky and Russell}(2008)}]{Kostelecky:2008ts}
\bibinfo{author}{\bibfnamefont{V.~A.} \bibnamefont{Kostelecky}}
  \bibnamefont{and} \bibinfo{author}{\bibfnamefont{N.}~\bibnamefont{Russell}}
  (\bibinfo{year}{2008}), \eprint{0801.0287}.

\bibitem[{\citenamefont{Lewis et~al.}(2000)\citenamefont{Lewis, Challinor, and
  Lasenby}}]{Lewis:1999bs}
\bibinfo{author}{\bibfnamefont{A.}~\bibnamefont{Lewis}},
  \bibinfo{author}{\bibfnamefont{A.}~\bibnamefont{Challinor}},
  \bibnamefont{and} \bibinfo{author}{\bibfnamefont{A.}~\bibnamefont{Lasenby}},
  \bibinfo{journal}{Astrophys. J.} \textbf{\bibinfo{volume}{538}},
  \bibinfo{pages}{473} (\bibinfo{year}{2000}), \eprint{astro-ph/9911177}.

\bibitem[{\citenamefont{{Nishizawa}}(2014)}]{2014PTEP.2014fB110N}
\bibinfo{author}{\bibfnamefont{A.~J.} \bibnamefont{{Nishizawa}}},
  \bibinfo{journal}{Progress of Theoretical and Experimental Physics}
  \textbf{\bibinfo{volume}{2014}}, \bibinfo{eid}{06B110}
  (\bibinfo{year}{2014}), \eprint{1404.5102}.

\bibitem[{\citenamefont{Lewis and Bridle}(2002)}]{Lewis:2002ah}
\bibinfo{author}{\bibfnamefont{A.}~\bibnamefont{Lewis}} \bibnamefont{and}
  \bibinfo{author}{\bibfnamefont{S.}~\bibnamefont{Bridle}},
  \bibinfo{journal}{Phys. Rev. D} \textbf{\bibinfo{volume}{66}},
  \bibinfo{pages}{103511} (\bibinfo{year}{2002}), \eprint{astro-ph/0205436}.

\bibitem[{\citenamefont{Aghanim et~al.}(2020{\natexlab{a}})}]{Planck:2019nip}
\bibinfo{author}{\bibfnamefont{N.}~\bibnamefont{Aghanim}} \bibnamefont{et~al.}
  (\bibinfo{collaboration}{Planck}), \bibinfo{journal}{Astron. Astrophys.}
  \textbf{\bibinfo{volume}{641}}, \bibinfo{pages}{A5}
  (\bibinfo{year}{2020}{\natexlab{a}}), \eprint{1907.12875}.

\bibitem[{\citenamefont{Aghanim et~al.}(2020{\natexlab{b}})}]{Planck:2018lbu}
\bibinfo{author}{\bibfnamefont{N.}~\bibnamefont{Aghanim}} \bibnamefont{et~al.}
  (\bibinfo{collaboration}{Planck}), \bibinfo{journal}{Astron. Astrophys.}
  \textbf{\bibinfo{volume}{641}}, \bibinfo{pages}{A8}
  (\bibinfo{year}{2020}{\natexlab{b}}), \eprint{1807.06210}.

\bibitem[{\citenamefont{Beutler et~al.}(2011)\citenamefont{Beutler, Blake,
  Colless, Jones, Staveley-Smith, Campbell, Parker, Saunders, and
  Watson}}]{Beutler:2011hx}
\bibinfo{author}{\bibfnamefont{F.}~\bibnamefont{Beutler}},
  \bibinfo{author}{\bibfnamefont{C.}~\bibnamefont{Blake}},
  \bibinfo{author}{\bibfnamefont{M.}~\bibnamefont{Colless}},
  \bibinfo{author}{\bibfnamefont{D.~H.} \bibnamefont{Jones}},
  \bibinfo{author}{\bibfnamefont{L.}~\bibnamefont{Staveley-Smith}},
  \bibinfo{author}{\bibfnamefont{L.}~\bibnamefont{Campbell}},
  \bibinfo{author}{\bibfnamefont{Q.}~\bibnamefont{Parker}},
  \bibinfo{author}{\bibfnamefont{W.}~\bibnamefont{Saunders}}, \bibnamefont{and}
  \bibinfo{author}{\bibfnamefont{F.}~\bibnamefont{Watson}},
  \bibinfo{journal}{Mon. Not. Roy. Astron. Soc.}
  \textbf{\bibinfo{volume}{416}}, \bibinfo{pages}{3017} (\bibinfo{year}{2011}),
  \eprint{1106.3366}.

\bibitem[{\citenamefont{Ross et~al.}(2015)\citenamefont{Ross, Samushia,
  Howlett, Percival, Burden, and Manera}}]{Ross:2014qpa}
\bibinfo{author}{\bibfnamefont{A.~J.} \bibnamefont{Ross}},
  \bibinfo{author}{\bibfnamefont{L.}~\bibnamefont{Samushia}},
  \bibinfo{author}{\bibfnamefont{C.}~\bibnamefont{Howlett}},
  \bibinfo{author}{\bibfnamefont{W.~J.} \bibnamefont{Percival}},
  \bibinfo{author}{\bibfnamefont{A.}~\bibnamefont{Burden}}, \bibnamefont{and}
  \bibinfo{author}{\bibfnamefont{M.}~\bibnamefont{Manera}},
  \bibinfo{journal}{Mon. Not. Roy. Astron. Soc.}
  \textbf{\bibinfo{volume}{449}}, \bibinfo{pages}{835} (\bibinfo{year}{2015}),
  \eprint{1409.3242}.

\bibitem[{\citenamefont{Alam et~al.}(2017)}]{BOSS:2016wmc}
\bibinfo{author}{\bibfnamefont{S.}~\bibnamefont{Alam}} \bibnamefont{et~al.}
  (\bibinfo{collaboration}{BOSS}), \bibinfo{journal}{Mon. Not. Roy. Astron.
  Soc.} \textbf{\bibinfo{volume}{470}}, \bibinfo{pages}{2617}
  (\bibinfo{year}{2017}), \eprint{1607.03155}.

\bibitem[{\citenamefont{Scolnic et~al.}(2018)}]{Scolnic:2017caz}
\bibinfo{author}{\bibfnamefont{D.~M.} \bibnamefont{Scolnic}}
  \bibnamefont{et~al.}, \bibinfo{journal}{Astrophys. J.}
  \textbf{\bibinfo{volume}{859}}, \bibinfo{pages}{101} (\bibinfo{year}{2018}),
  \eprint{1710.00845}.

\bibitem[{\citenamefont{Jimenez and Loeb}(2002)}]{Jimenez:2001gg}
\bibinfo{author}{\bibfnamefont{R.}~\bibnamefont{Jimenez}} \bibnamefont{and}
  \bibinfo{author}{\bibfnamefont{A.}~\bibnamefont{Loeb}},
  \bibinfo{journal}{Astrophys. J.} \textbf{\bibinfo{volume}{573}},
  \bibinfo{pages}{37} (\bibinfo{year}{2002}), \eprint{astro-ph/0106145}.

\bibitem[{\citenamefont{Stern et~al.}(2010)\citenamefont{Stern, Jimenez, Verde,
  Kamionkowski, and Stanford}}]{Stern:2009ep}
\bibinfo{author}{\bibfnamefont{D.}~\bibnamefont{Stern}},
  \bibinfo{author}{\bibfnamefont{R.}~\bibnamefont{Jimenez}},
  \bibinfo{author}{\bibfnamefont{L.}~\bibnamefont{Verde}},
  \bibinfo{author}{\bibfnamefont{M.}~\bibnamefont{Kamionkowski}},
  \bibnamefont{and} \bibinfo{author}{\bibfnamefont{S.~A.}
  \bibnamefont{Stanford}}, \bibinfo{journal}{JCAP}
  \textbf{\bibinfo{volume}{02}}, \bibinfo{pages}{008} (\bibinfo{year}{2010}),
  \eprint{0907.3149}.

\bibitem[{\citenamefont{Moresco}(2015)}]{Moresco:2015cya}
\bibinfo{author}{\bibfnamefont{M.}~\bibnamefont{Moresco}},
  \bibinfo{journal}{Mon. Not. Roy. Astron. Soc.}
  \textbf{\bibinfo{volume}{450}}, \bibinfo{pages}{L16} (\bibinfo{year}{2015}),
  \eprint{1503.01116}.

\bibitem[{\citenamefont{Zhang et~al.}(2014)\citenamefont{Zhang, Zhang, Yuan,
  Zhang, and Sun}}]{Zhang:2012mp}
\bibinfo{author}{\bibfnamefont{C.}~\bibnamefont{Zhang}},
  \bibinfo{author}{\bibfnamefont{H.}~\bibnamefont{Zhang}},
  \bibinfo{author}{\bibfnamefont{S.}~\bibnamefont{Yuan}},
  \bibinfo{author}{\bibfnamefont{T.-J.} \bibnamefont{Zhang}}, \bibnamefont{and}
  \bibinfo{author}{\bibfnamefont{Y.-C.} \bibnamefont{Sun}},
  \bibinfo{journal}{Res. Astron. Astrophys.} \textbf{\bibinfo{volume}{14}},
  \bibinfo{pages}{1221} (\bibinfo{year}{2014}), \eprint{1207.4541}.

\bibitem[{\citenamefont{Moresco et~al.}(2016)\citenamefont{Moresco, Pozzetti,
  Cimatti, Jimenez, Maraston, Verde, Thomas, Citro, Tojeiro, and
  Wilkinson}}]{Moresco:2016mzx}
\bibinfo{author}{\bibfnamefont{M.}~\bibnamefont{Moresco}},
  \bibinfo{author}{\bibfnamefont{L.}~\bibnamefont{Pozzetti}},
  \bibinfo{author}{\bibfnamefont{A.}~\bibnamefont{Cimatti}},
  \bibinfo{author}{\bibfnamefont{R.}~\bibnamefont{Jimenez}},
  \bibinfo{author}{\bibfnamefont{C.}~\bibnamefont{Maraston}},
  \bibinfo{author}{\bibfnamefont{L.}~\bibnamefont{Verde}},
  \bibinfo{author}{\bibfnamefont{D.}~\bibnamefont{Thomas}},
  \bibinfo{author}{\bibfnamefont{A.}~\bibnamefont{Citro}},
  \bibinfo{author}{\bibfnamefont{R.}~\bibnamefont{Tojeiro}}, \bibnamefont{and}
  \bibinfo{author}{\bibfnamefont{D.}~\bibnamefont{Wilkinson}},
  \bibinfo{journal}{JCAP} \textbf{\bibinfo{volume}{05}}, \bibinfo{pages}{014}
  (\bibinfo{year}{2016}), \eprint{1601.01701}.

\bibitem[{\citenamefont{Ratsimbazafy et~al.}(2017)\citenamefont{Ratsimbazafy,
  Loubser, Crawford, Cress, Bassett, Nichol, and
  V\"ais\"anen}}]{Ratsimbazafy:2017vga}
\bibinfo{author}{\bibfnamefont{A.~L.} \bibnamefont{Ratsimbazafy}},
  \bibinfo{author}{\bibfnamefont{S.~I.} \bibnamefont{Loubser}},
  \bibinfo{author}{\bibfnamefont{S.~M.} \bibnamefont{Crawford}},
  \bibinfo{author}{\bibfnamefont{C.~M.} \bibnamefont{Cress}},
  \bibinfo{author}{\bibfnamefont{B.~A.} \bibnamefont{Bassett}},
  \bibinfo{author}{\bibfnamefont{R.~C.} \bibnamefont{Nichol}},
  \bibnamefont{and}
  \bibinfo{author}{\bibfnamefont{P.}~\bibnamefont{V\"ais\"anen}},
  \bibinfo{journal}{Mon. Not. Roy. Astron. Soc.}
  \textbf{\bibinfo{volume}{467}}, \bibinfo{pages}{3239} (\bibinfo{year}{2017}),
  \eprint{1702.00418}.

\bibitem[{\citenamefont{Riess et~al.}(2016)}]{Riess:2016jrr}
\bibinfo{author}{\bibfnamefont{A.~G.} \bibnamefont{Riess}}
  \bibnamefont{et~al.}, \bibinfo{journal}{Astrophys. J.}
  \textbf{\bibinfo{volume}{826}}, \bibinfo{pages}{56} (\bibinfo{year}{2016}),
  \eprint{1604.01424}.

\bibitem[{\citenamefont{Camarena and Marra}(2021)}]{Camarena:2021jlr}
\bibinfo{author}{\bibfnamefont{D.}~\bibnamefont{Camarena}} \bibnamefont{and}
  \bibinfo{author}{\bibfnamefont{V.}~\bibnamefont{Marra}},
  \bibinfo{journal}{Mon. Not. Roy. Astron. Soc.}
  \textbf{\bibinfo{volume}{504}}, \bibinfo{pages}{5164} (\bibinfo{year}{2021}),
  \eprint{2101.08641}.

\bibitem[{\citenamefont{Benetti et~al.}(2017)\citenamefont{Benetti, Graef, and
  Alcaniz}}]{Benetti:2017gvm}
\bibinfo{author}{\bibfnamefont{M.}~\bibnamefont{Benetti}},
  \bibinfo{author}{\bibfnamefont{L.~L.} \bibnamefont{Graef}}, \bibnamefont{and}
  \bibinfo{author}{\bibfnamefont{J.~S.} \bibnamefont{Alcaniz}},
  \bibinfo{journal}{JCAP} \textbf{\bibinfo{volume}{04}}, \bibinfo{pages}{003}
  (\bibinfo{year}{2017}), \eprint{1702.06509}.

\bibitem[{\citenamefont{Benetti et~al.}(2018)\citenamefont{Benetti, Graef, and
  Alcaniz}}]{Benetti:2017juy}
\bibinfo{author}{\bibfnamefont{M.}~\bibnamefont{Benetti}},
  \bibinfo{author}{\bibfnamefont{L.~L.} \bibnamefont{Graef}}, \bibnamefont{and}
  \bibinfo{author}{\bibfnamefont{J.~S.} \bibnamefont{Alcaniz}},
  \bibinfo{journal}{JCAP} \textbf{\bibinfo{volume}{07}}, \bibinfo{pages}{066}
  (\bibinfo{year}{2018}), \eprint{1712.00677}.

\bibitem[{\citenamefont{Efstathiou and Gratton}(2020)}]{Efstathiou:2020wem}
\bibinfo{author}{\bibfnamefont{G.}~\bibnamefont{Efstathiou}} \bibnamefont{and}
  \bibinfo{author}{\bibfnamefont{S.}~\bibnamefont{Gratton}}
  (\bibinfo{year}{2020}), \eprint{2002.06892}.

\bibitem[{\citenamefont{Gonzalez et~al.}(2021)\citenamefont{Gonzalez, Benetti,
  von Marttens, and Alcaniz}}]{Gonzalez:2021ojp}
\bibinfo{author}{\bibfnamefont{J.~E.} \bibnamefont{Gonzalez}},
  \bibinfo{author}{\bibfnamefont{M.}~\bibnamefont{Benetti}},
  \bibinfo{author}{\bibfnamefont{R.}~\bibnamefont{von Marttens}},
  \bibnamefont{and} \bibinfo{author}{\bibfnamefont{J.}~\bibnamefont{Alcaniz}},
  \bibinfo{journal}{JCAP} \textbf{\bibinfo{volume}{11}}, \bibinfo{pages}{060}
  (\bibinfo{year}{2021}), \eprint{2104.13455}.

\bibitem[{\citenamefont{Vagnozzi et~al.}(2021)\citenamefont{Vagnozzi, Loeb, and
  Moresco}}]{vagnozzi:2020dfn}
\bibinfo{author}{\bibfnamefont{S.}~\bibnamefont{Vagnozzi}},
  \bibinfo{author}{\bibfnamefont{A.}~\bibnamefont{Loeb}}, \bibnamefont{and}
  \bibinfo{author}{\bibfnamefont{M.}~\bibnamefont{Moresco}},
  \bibinfo{journal}{Astrophys. J.} \textbf{\bibinfo{volume}{908}},
  \bibinfo{pages}{84} (\bibinfo{year}{2021}), \eprint{2011.11645}.

\bibitem[{\citenamefont{Moresco et~al.}(2022)}]{Moresco:2022phi}
\bibinfo{author}{\bibfnamefont{M.}~\bibnamefont{Moresco}} \bibnamefont{et~al.}
  (\bibinfo{year}{2022}), \eprint{2201.07241}.

\bibitem[{\citenamefont{{Planck Collaboration}
  et~al.}(2016)\citenamefont{{Planck Collaboration}, {Aghanim}, {Arnaud},
  {Ashdown}, {Aumont}, {Baccigalupi}, {Banday}, {Barreiro}, {Bartlett},
  {Bartolo} et~al.}}]{2016A&A...594A..11P}
\bibinfo{author}{\bibnamefont{{Planck Collaboration}}},
  \bibinfo{author}{\bibfnamefont{N.}~\bibnamefont{{Aghanim}}},
  \bibinfo{author}{\bibfnamefont{M.}~\bibnamefont{{Arnaud}}},
  \bibinfo{author}{\bibfnamefont{M.}~\bibnamefont{{Ashdown}}},
  \bibinfo{author}{\bibfnamefont{J.}~\bibnamefont{{Aumont}}},
  \bibinfo{author}{\bibfnamefont{C.}~\bibnamefont{{Baccigalupi}}},
  \bibinfo{author}{\bibfnamefont{A.~J.} \bibnamefont{{Banday}}},
  \bibinfo{author}{\bibfnamefont{R.~B.} \bibnamefont{{Barreiro}}},
  \bibinfo{author}{\bibfnamefont{J.~G.} \bibnamefont{{Bartlett}}},
  \bibinfo{author}{\bibfnamefont{N.}~\bibnamefont{{Bartolo}}},
  \bibnamefont{et~al.}, \bibinfo{journal}{Astronomy and Astrophysics}
  \textbf{\bibinfo{volume}{594}}, \bibinfo{eid}{A11} (\bibinfo{year}{2016}),
  \eprint{1507.02704}.

\bibitem[{\citenamefont{{Liddle}}(2007)}]{DIC}
\bibinfo{author}{\bibfnamefont{A.~R.} \bibnamefont{{Liddle}}},
  \bibinfo{journal}{Monthly Notices Royal Astronomical Society}
  \textbf{\bibinfo{volume}{377}}, \bibinfo{pages}{L74} (\bibinfo{year}{2007}),
  \eprint{astro-ph/0701113}.

\bibitem[{\citenamefont{Kunz et~al.}(2006)\citenamefont{Kunz, Trotta, and
  Parkinson}}]{Kunz:2006mc}
\bibinfo{author}{\bibfnamefont{M.}~\bibnamefont{Kunz}},
  \bibinfo{author}{\bibfnamefont{R.}~\bibnamefont{Trotta}}, \bibnamefont{and}
  \bibinfo{author}{\bibfnamefont{D.}~\bibnamefont{Parkinson}},
  \bibinfo{journal}{Phys. Rev.} \textbf{\bibinfo{volume}{D74}},
  \bibinfo{pages}{023503} (\bibinfo{year}{2006}), \eprint{astro-ph/0602378}.

\bibitem[{\citenamefont{Trotta}(2008)}]{Trotta:2008qt}
\bibinfo{author}{\bibfnamefont{R.}~\bibnamefont{Trotta}},
  \bibinfo{journal}{Contemp. Phys.} \textbf{\bibinfo{volume}{49}},
  \bibinfo{pages}{71} (\bibinfo{year}{2008}), \eprint{0803.4089}.

\bibitem[{\citenamefont{Spiegelhalter et~al.}(2002)\citenamefont{Spiegelhalter,
  Best, Carlin, and van~der Linde}}]{Spiegelhalter:2002yvw}
\bibinfo{author}{\bibfnamefont{D.~J.} \bibnamefont{Spiegelhalter}},
  \bibinfo{author}{\bibfnamefont{N.~G.} \bibnamefont{Best}},
  \bibinfo{author}{\bibfnamefont{B.~P.} \bibnamefont{Carlin}},
  \bibnamefont{and} \bibinfo{author}{\bibfnamefont{A.}~\bibnamefont{van~der
  Linde}}, \bibinfo{journal}{J. Roy. Statist. Soc. B}
  \textbf{\bibinfo{volume}{64}}, \bibinfo{pages}{583} (\bibinfo{year}{2002}).

\bibitem[{\citenamefont{Santos~da Costa et~al.}(2021)\citenamefont{Santos~da
  Costa, Benetti, Neves, Brito, Silva, and Alcaniz}}]{SantosdaCosta:2020dyl}
\bibinfo{author}{\bibfnamefont{S.}~\bibnamefont{Santos~da Costa}},
  \bibinfo{author}{\bibfnamefont{M.}~\bibnamefont{Benetti}},
  \bibinfo{author}{\bibfnamefont{R.~M.~P.} \bibnamefont{Neves}},
  \bibinfo{author}{\bibfnamefont{F.~A.} \bibnamefont{Brito}},
  \bibinfo{author}{\bibfnamefont{R.}~\bibnamefont{Silva}}, \bibnamefont{and}
  \bibinfo{author}{\bibfnamefont{J.~S.} \bibnamefont{Alcaniz}},
  \bibinfo{journal}{Eur. Phys. J. Plus} \textbf{\bibinfo{volume}{136}},
  \bibinfo{pages}{84} (\bibinfo{year}{2021}), \eprint{2007.09211}.

\bibitem[{\citenamefont{Frusciante and Benetti}(2021)}]{Frusciante:2020gkx}
\bibinfo{author}{\bibfnamefont{N.}~\bibnamefont{Frusciante}} \bibnamefont{and}
  \bibinfo{author}{\bibfnamefont{M.}~\bibnamefont{Benetti}},
  \bibinfo{journal}{Phys. Rev. D} \textbf{\bibinfo{volume}{103}},
  \bibinfo{pages}{104060} (\bibinfo{year}{2021}), \eprint{2005.14705}.

\bibitem[{\citenamefont{Winkler et~al.}(2020)\citenamefont{Winkler, Gerbino,
  and Benetti}}]{Winkler:2019hkh}
\bibinfo{author}{\bibfnamefont{M.~W.} \bibnamefont{Winkler}},
  \bibinfo{author}{\bibfnamefont{M.}~\bibnamefont{Gerbino}}, \bibnamefont{and}
  \bibinfo{author}{\bibfnamefont{M.}~\bibnamefont{Benetti}},
  \bibinfo{journal}{Phys. Rev. D} \textbf{\bibinfo{volume}{101}},
  \bibinfo{pages}{083525} (\bibinfo{year}{2020}), \eprint{1911.11148}.

\bibitem[{\citenamefont{Kass and Raftery}(1995)}]{Kass:1995loi}
\bibinfo{author}{\bibfnamefont{R.~E.} \bibnamefont{Kass}} \bibnamefont{and}
  \bibinfo{author}{\bibfnamefont{A.~E.} \bibnamefont{Raftery}},
  \bibinfo{journal}{J. Am. Statist. Assoc.} \textbf{\bibinfo{volume}{90}},
  \bibinfo{pages}{773} (\bibinfo{year}{1995}).

\end{thebibliography}
\appendix
\section{Effects of the friction driven force in Black Holes}
\label{BH_effects}
In this appendix we discuss the effects of the force driven by the friction that arises when the effects of granularity of space-times are considered in black holes. 
 In this regard, general considerations  led us to   postulate a rotational friction term of the form: 
   \be\label{spinny}
    u^{\mu}\nabla_{\mu} s^{\nu}=\overline\alpha_{\rm bh} \frac{M}{m^2_{\rm p}}\, {\rm sign}(s\cdot \xi)\tilde { \mathbf R}\, (s\cdot s)\, u^{\nu}- \overline\beta_{\rm bh} \frac{M}{m^2_{\rm p}}\, \tilde { \mathbf R}_{\rm BH}\, s^{\nu},\ee
    where $M$ is the black hole's mass,  $\overline\alpha_{\rm bh}$   is connected to  the translational friction term analogous to the  one  affecting particles  (as in eq.  \ref{modimodi})   and  the second term  takes into account that,  in contrast to elementary particles, the   spin  of the black hole   might  change  not only in orientation  but also in  magnitude.   Thus  $\overline\beta_{\rm bh}$ is a new dimensionless  parameter  characterizing the  ``intrinsic" spin diffusion term. 
In \cite{Perez:2019gyd}  two   options  for the   natural order of magnitude of these effective parameters  were  considered, involving a  {\it ``high"}  and {\it ``low"}  suppression levels.
In order to  asses  the   magnitude of  the effect  under consideration,  we compare the order of magnitude  or  the   resulting     anomalous force, which   we denote by  $F$,   with that of the {\it  standard}  gravitational force  between  two   black holes  at the moment of closest approach in  a black hole coalescence event  which we   denote by $ F_{\rm BHC}$. The estimate of  $ F_{\rm BHC}$ is  based on the   consideration  of   two equal mass BH's and the use  of  the Newtonian expression,   evaluated  when the separation  between  both is of the order of the Schwarzschild radius,  and the anomalous  force  is  estimated for the  case  the two    black holes are extremal so $s = (M/m_p)^2$ (the magnitude  of the  black hole's  spin measured in  natural units).  The order of  magnitude  estimate for such quantity turns out to be  gives  $ |F_{\rm BHC}| \sim  GM^2/(2GM)^2 = m^2_p/4$.  It is  worth  emphasizing that while  $F_{\rm BHC}$ is taken to  represent the {\it standard gravitational interaction}, the  hypothetical anomalous force   $F$ we  are considering would  be also, ultimately, of gravitational origin,  although, in this  case,   tied intrinsically  with the granular aspects,  of space-time  which we take to characterize its underlying  fundamental  quantum gravity origin. As  an analogy one might consider comparing  an electromagnetic  force of between two magnets  with the friction force that affects their motion on  a surface, which  is,  of course,    also  of  electromagnetic  origin.
The  results are the following \cite{Perez:2019gyd}:    in the case   involving a    high suppression  level   

\be\label{GW1} \boxed{\overline\alpha_{\rm bh}=\overline\alpha_{\rm bh}^{01} {\frac{m_{\rm p}}{M}}\ \ \ \Longrightarrow \ \ \ \left|\frac{F}{F_{\rm BHC}}\right|\le
4\overline\alpha^{01}_{\rm bh} 10^{-6} \left(\frac{M}{M_{\odot}}\right)^3},\ee
and for the  second possibility involving   a  low  level  of suppression
\be\label{GW2}\boxed{ \overline\alpha_{\rm bh}=\overline\alpha_{\rm bh}^{02} \sqrt{\frac{m_{\rm p}}{M}}\ \  \Longrightarrow \ \ \left|\frac{F}{F_{\rm BHC}}\right|\le
4\overline\alpha^{02}_{\rm bh} 10^{13} \left(\frac{M}{M_{\odot}}\right)^{\frac72}}.\ee


These two possibilities  might  be  seen as derived from a $1/\sqrt{N}$ suppression of some stochastic origin with the number of `area quanta' $N\approx  M^2/m^2_{\rm p}$ or the  number of `energy-quanta' $N\approx M/m_{\rm p}$ involved respectively.

The quantity $\tilde { \mathbf R}_{\rm BH}$  represents an appropriate  measure of the  mean local curvature  in the  surroundings  of  the black hole.  Such  quantity could be  for instance   something like an  ``averaged value"  of the local Kretschman  scalar $\sqrt{ R_{abcd}R^{abcd}} $,  in the  region  occupied by the black hole.  For simplicity in \cite{Perez:2020cwa}  we   explored the  simple  case in which that  quantity  is
  taken to be a  simple estimate of  curvature $\tilde { \mathbf R}_{\rm BH}=1/M^2$. Once this choice is made,   and   taking   natural order of magnitude estimates for  the parameters $\overline\alpha_{\rm bh}$    and  $\overline\beta_{\rm bh}$ the reminder of the analysis  would  in  principle be quite direct  but    heavily dependent on  aspects of   astrophysics  and  cosmology on which our  knowledge  is   still  quite  incipient. 

\end{document}